\documentclass[useAMS,usenatbib]{mnras}

\usepackage[authoryear]{natbib}

\usepackage{txfonts}
\usepackage{graphicx}
\usepackage{pdflscape}
\usepackage{threeparttable}
\def\mum{\,$\mu$m}
\def\deg{^{\circ}}
\def\sun{$_{\odot}$}
\def\Hyp{\textit{Hyper}}

\def\req{\,$R_{eq}$}
\def\12co{$^{12}$CO}
\def\co13{$^{13}$CO}
\def\NH2{N$_{\mathrm{H}_{2}}$}
\def\n2h{N$_{2}$H$^{+}$}
\def\d2n{N$_{2}$D$^{+}$}
\def\nh3{NH$_{3}$}
\def\hco{HCO$^{+}$}
\def\h13co{H$^{13}$CO}
\def\h20{H$_{2}$O}

\def\avir{$\alpha_{vir}$}
\def\0avir{$\alpha_{0}$}

\def\tex{$T_{ex}$}

\def\a_eff{$\tilde{\alpha}_{eff}$}

\def\sn{$S/N$}

\title{Multi-scale dynamics in star-forming regions: the interplay between gravity and turbulence}

\author[A. Traficante]{A. Traficante$^{1,2}$\thanks{e-mail:alessio.traficante@iaps.inaf.it}, G.A. Fuller$^{2}$, A. Duarte-Cabral$^{3}$, D. Elia$^{1}$, M. H. Heyer$^{4}$, S. Molinari$^{1}$,
\newauthor N. Peretto$^{3}$, E. Schisano$^{1}$ \\
$^{1}$IAPS - INAF, via Fosso del Cavaliere, 100, I-00133 Roma, Italy \\
$^{2}$Jodrell Bank Centre for Astrophysics, Department of Physics and Astronomy, The University of Manchester, Oxford Road, Manchester M13 9PL, UK \\
$^{3}$School of Physics and Astronomy, Cardiff University, Queens Buildings, The Parade, Cardiff CF24 3AA, UK \\
$^{4}$Department of Astronomy, University of Massachusetts, Amherst, MA 01003, USA}

\newcommand{\GG}[1]{}

\begin{document}
\maketitle

\label{firstpage}

\begin{abstract}
In this work we investigate the interplay between gravity and turbulence at different spatial scales and in different density regimes. We analyze a sample of 70\mum\ quiet clumps that are divided into three surface density bins and we compare the dynamics of each group with the dynamics of their respective filaments. The densest clumps form within the densest filaments on average, and they have the highest value of the velocity dispersion. The kinetic energy is transferred from the filaments down to the clumps most likely through a turbulent cascade, but we identify a critical value of the surface density, $\Sigma\simeq0.1$ g cm$^{-2}$, above which the dynamics changes from being mostly turbulent-driven to mostly gravity-driven. The scenario we obtain from our data is a continuous interplay between turbulence and gravity, where the former creates structures at all scales and the latter takes the lead when the critical surface density threshold is reached. In the densest filaments this transition can occur at the parsec, or even larger scales, leading to a global collapse of the whole region and most likely to the formation of the massive objects.
\end{abstract}


\section{Introduction}
The massive star formation involves a hierarchical, multi-scale process that starts in giant molecular clouds (GMCs), objects with size up to tens of parsecs \citep{Solomon87,Roman-Duval10}. Within GMCs, the formation of stars begins preferably across elongated features called filaments, which are ubiquitous in the Galaxy \citep{Molinari10_PASP,Andre14,Schisano14,Schisano19,Li16_fil}. Along these filaments, parsec-scale massive and dense clumps develop average surface density larger than 0.05-0.1 g cm$^{-2}$ \citep{Urquhart14,Traficante15a,Svoboda16}, which lead to the formation of the first protostellar cores \citep{Zhang15,Csengeri17} that will finally produce a cluster of stars \citep[e.g.][]{Lada03}, including the most massive ones. 

These different structures are identified at the various spatial scales, but it is not yet clearly understood if (and eventually how) they are dynamically coupled. Turbulent accretion models \citep[e.g.][]{McKee03} predict that the collapse occurs \textit{locally} and in a similar fashion to what happens in low-mass regions: massive clumps are relatively isolated from the rest of the clouds. They are maintained close to the virial equilibrium thanks to the high pressure sustained by the supersonic non-thermal motions driven by the turbulence of the local interstellar-medium. Other models predict  that the whole cloud is in hierarchical, \textit{global} collapse and the supersonic, non-thermal velocity dispersions reflect infalling motions at all spatial scales \citep{Vazquez-Semadeni09,Vazquez-Semadeni19,Ballesteros-Paredes11}.

Observations of massive, filamentary infrared dark clouds (IRDCs) support the global collapse mechanism \citep[e.g.][]{Peretto13}, with evidence that large-scale motions driven by self-gravity occur simultaneously and are coupled to the local motions, with a net increase of supersonic motions along the embedded clumps due to the longitudinal collapse of the whole cloud, like in SDC13 \citep{Peretto14}, or along the junctions of filaments, where dense clumps are forming, like in Monoceros OB1 \citep{Montillaud19}. The gravitational fragmentation along a filament may also be regulated by the turbulence that acts at all spatial scales, as suggested from the dynamics observed in G26 \citep{Liu18}. In contrast, a cloud such as G035 shows that the dynamics of the filaments and the embedded clumps are different and they may be considered as independent structures, with non-thermal motions that are only mildly supersonic \citep{Henshaw14}.

These few examples illustrate the need to distinguish between non-thermal motions driven by gravity or regulated by turbulence in massive star forming regions \citep{Motte18,Krumholz18}, as well as the need to determine whether the dynamics of the collapse in the high-mass regime differs or not from the low-mass counterpart.

With large surveys of the Galactic Plane that observed the cold dust continuum emission of thousands of star-forming regions (e.g. Hi-GAL, \citealt[][]{Molinari10_PASP}; ATLASGAL, \citealt[][]{Schuller09}; BGPS, \citealt[][]{Aguirre10}) and the relatively low density gas associated with molecular clouds and filaments, traced with the various CO isotopologues (e.g. \co13 $(1-0)$, GRS, \citealt{Jackson06}; \12co\ $(1-0)$ and \co13\ $(1-0)$, FUGIN, \citealt{Umemoto17}; \co13 $(2-1)$ and C$^{18}$O $(2-1)$, SEDIGISM, \citealt{Schuller17}; \co13 $(3-2)$ and C$^{18}$O $(3-2)$, CHIMPS \citealt{Rigby19}) we now have a rich database to statistically analyze the dynamics of star-forming regions in different environments. These catalogues, combined with observations of high-density tracers to identify the kinematics within the embedded clumps, enable a multi-scale study of the kinematics and dynamics in different mass and density regimes.

In this work we present new data obtained with the IRAM 30m telescope relative to 13 clumps selected from the Hi-GAL sample to be 70\mum\ quiet and in a range of low- to intermediate-surface densities, together with 16 high-mass 70\mum\ quiet clumps already studied in \citet{Traficante17_PI}. We combine these data with the properties of their parent filaments extracted from the newly published catalogue of filaments extracted from the Hi-GAL survey \citep{Schisano19}, for which we derived the dynamics using FUGIN and GRS data. We aim at studying the physical and kinematic properties of star-forming regions in different density regimes and at various spatial scales to investigate if there is a critical density regime above which gravity starts to dominate over turbulence, and at which scales the collapse can be considered global. The paper is organized as follow: in Section \ref{sec:observations} we describe the datasets we use to determine the properties of the clumps and parent filaments; in Section \ref{sec:results} we divide the clumps in three different density regimes and we discuss the physical properties and the dynamics of the clumps and the parent filaments in these different regimes; in Section \ref{sec:clumps_clouds_dynamics} we combine the dynamics of the clumps and the parent filaments, for a multi-scale analysis of these star-forming regions; in Section \ref{sec:non_thermal_motions_filaments} we discuss the scenario that emerges from our observations and, finally, in Section \ref{sec:discussion_conclusions} we present our summary.

\section{Observations}\label{sec:observations}

\subsection{Clump data}
The clumps used in this work combine two samples of so-called 70\mum\ quiet clumps, due to the lack (or extremely faint) emission at this wavelength, which is attributed to the lack of presence of an evolved protostar \citep{Dunham08}. We focus on this extremely young stage of the evolution to assure that any feedback from the presence of an embedded protostar is minimal. The first sample is a group of 16 massive massive and dense ($\Sigma\geq 0.05$ g cm$^{-2}$) clumps presented in \citep{Traficante17_PI}, and it is combined with a new sample of twenty-three 70\mum\ quiet clumps extracted from the catalogue of candidates starless sources embedded in IRDCs \citet{Traficante15a} and selected to cover a lower range of surface densities, with $0.005\leq\Sigma\leq0.05$ g cm$^{-2}$. The physical properties of the clumps have been re-derived from the \textit{Herschel} Hi-GAL survey data \citep{Molinari10_PASP} and including also the emission at 870\mum\ and 1.1 mm from the ATLASGAL \citep{Schuller09} and the Bolocam Galactic plane survey \citep[BGPS,][]{Aguirre10} respectively.

The gas kinematics of each clump have been obtained with 3mm observations centered on the \n2h\ $(1-0)$, \hco\ $(1-0)$ and HNC $(1-0)$ lines, acquired at the IRAM 30m telescope in December 2015 under the project 133-15. The sources were observed in On-The-Fly mode covering a $2\arcmin\times2\arcmin$ region centered on the peak of the clump, identified in the 250\mum\ Hi-GAL map. The spatial resolution is $\simeq28\arcsec$. The system temperature of the observations was in the range $85\leq T_{sys}\leq125$ K, depending on the source. We used the FTS50 backend to obtain a spectral resolution of 0.2 km s$^{-1}$, enough to resolve the hyperfine components of the \n2h\ $(1-0)$ emission line. We achieved a noise per channel per pixel $\leq0.16$ K in $T_{mb}$, assuming a beam efficiency at our frequencies of 0.78 \citep{Rygl13}. We averaged the emission across each clump and considered only clumps with a signal-to-noise ratio (\sn) of the \n2h\ $(1-0)$ average spectra above 5$\sigma$, where the noise level in each clump is determined as the standard deviation of the emission-free channels using the GILDAS/CLASS\footnote{http://www.iram.fr/IRAMFR/GILDAS} software package. 

Four sources out of 23 have no detected emission in the \n2h\ $(1-0)$ at the sensitivity of our observations, while three more sources have a \sn$<5$ and are therefore removed from the sample studied here. We also removed from the analysis the clumps 18.287-0.256, 33.332-0.531 and 49.57-0.192, since the \n2h\ $(1-0)$ spectrum showed two components along the line of sight that cannot be disentangled in the dust continuum data. 

\begin{figure*}
\centering
\includegraphics[width=4.5cm]{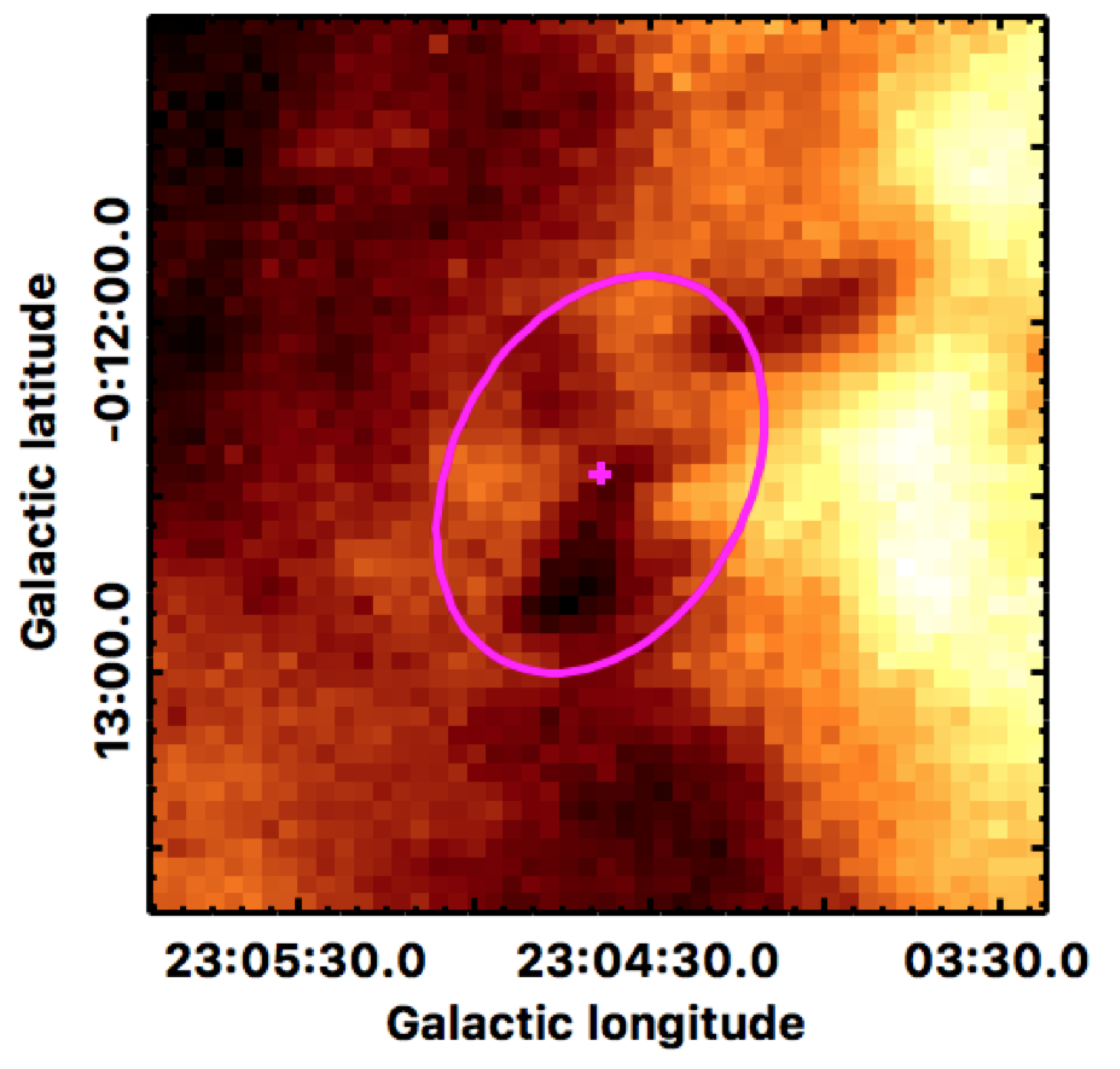}\qquad \qquad
\includegraphics[width=4.5cm]{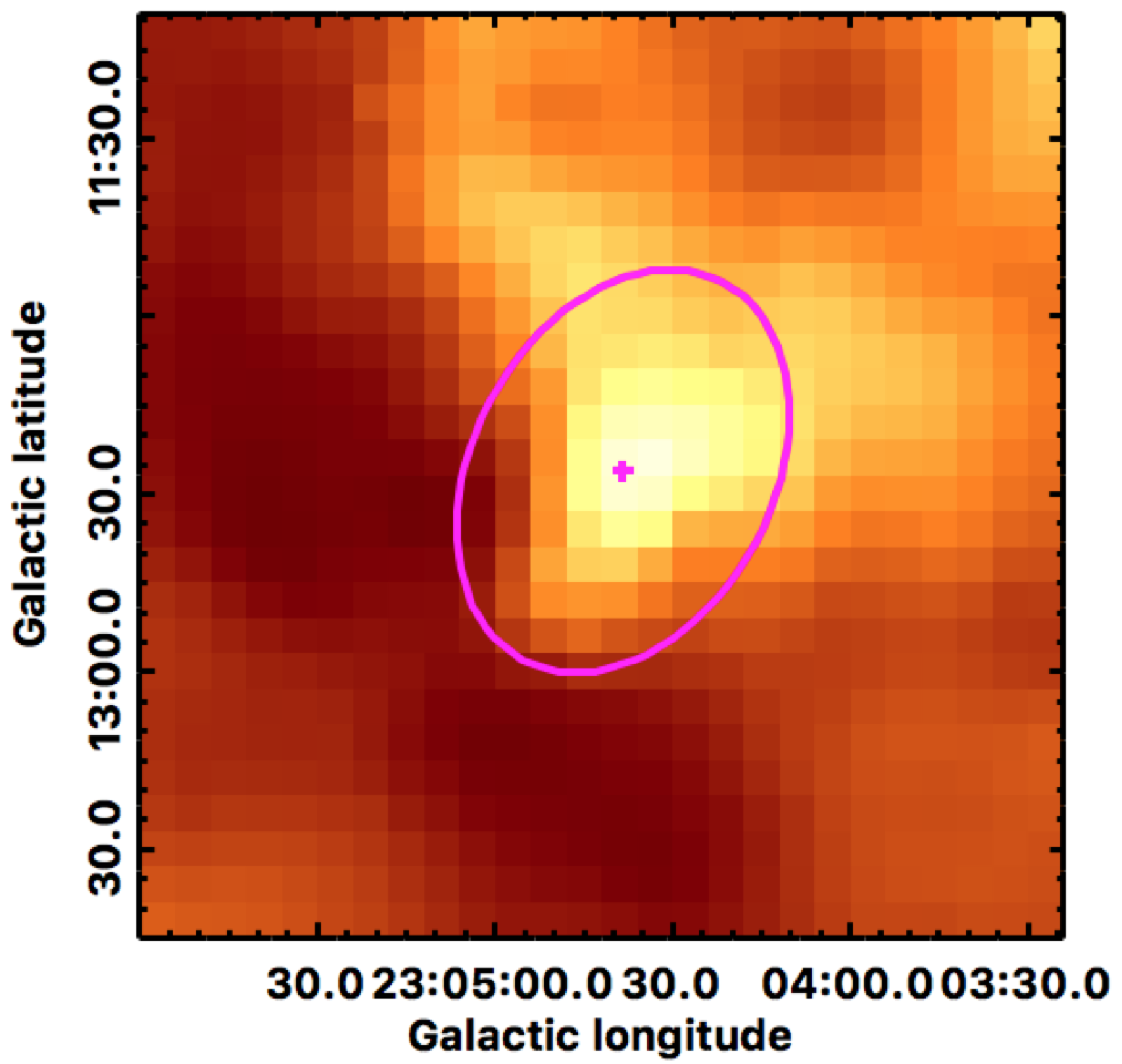}\qquad \qquad
\includegraphics[width=6cm]{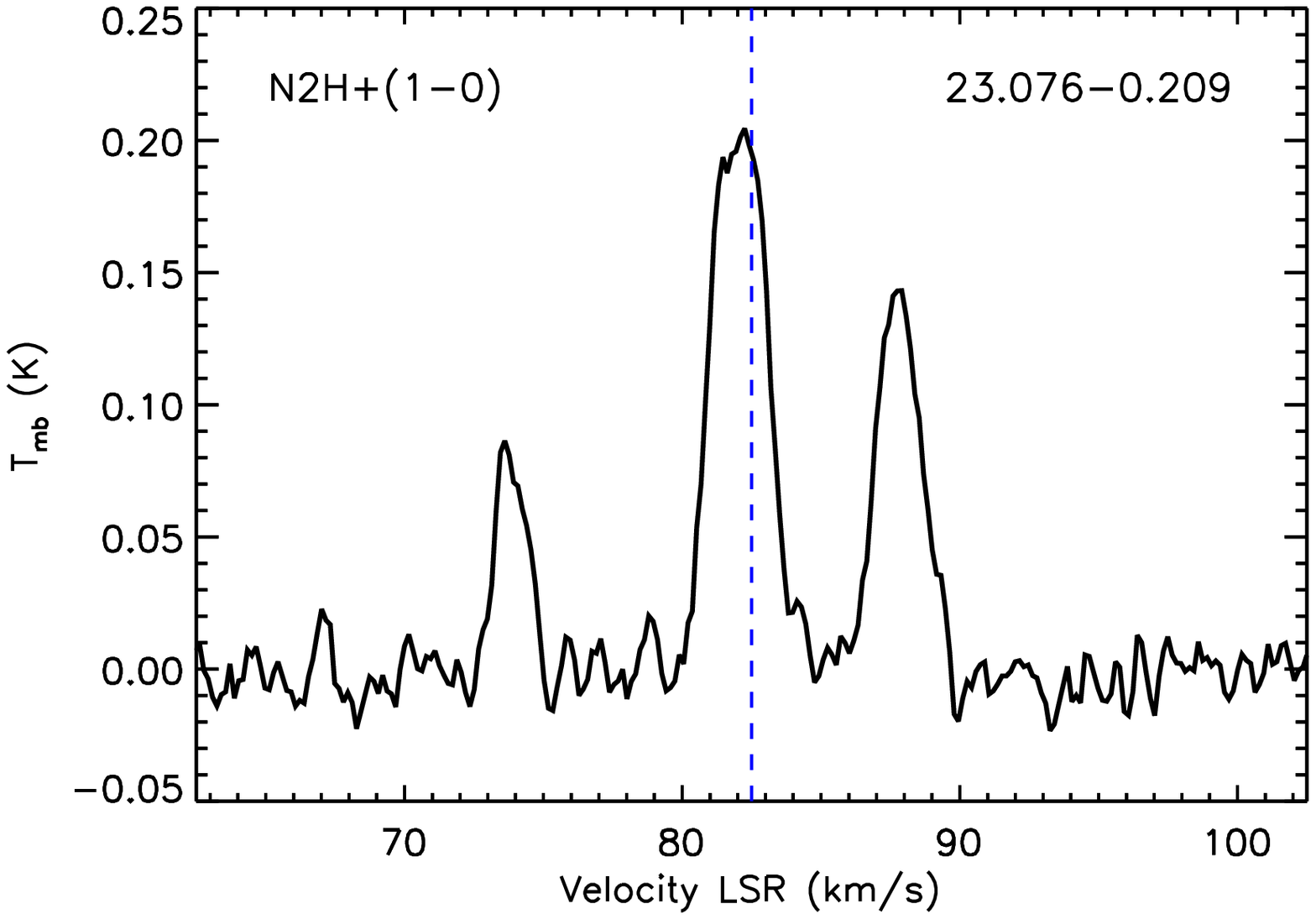}
\caption{The 70\mum\ quiet clump 23.076-0.209, belonging to the $\Sigma_{low}$ group, as seen in Hi-GAL at 70\mum\ (left panel) and at 250\mum\ (central panel). The magenta cross represents the 250\mum\ centroid and the magenta ellipse is the result of the 2d-Gaussian fitting performed with \textit{Hyper}. In the right panel, the \n2h\ $(1-0)$ emission averaged across the clump. The blue-dotted vertical line is in correspondence of the central velocity of the clump obtained with an hyperfine fitting of the \n2h\ spectrum done with CLASS using the standard routines.}    
\label{fig:23076-0209}
\end{figure*}

We ended-up with a total of 13 out of 23 sources with defined physical and kinematic properties, which, combined with the sample of 16 massive 70\mum\ quiet clumps of \citet{Traficante17_PI}, provides us with a total of 29 sources covering a range of surface densities $0.005\leq\Sigma\leq0.25$ g cm$^{-2}$. The example of the new 70\mum\ quiet sources 23.076-0.209 as seen at 70\mum\ and 250\mum\ and of its \n2h\ $(1-0)$ spectrum are shown in Figure \ref{fig:23076-0209}.

\subsection{Filament data}
For each clump we identified the parent filament from the catalogue of filaments in the Galactic plane extracted from the Hi-GAL survey \citep{Schisano19} through the algorithm described in \citet{Schisano14}. This algorithm extracts all the elongated structures in the Hi-GAL column density maps that have a higher contrast with respect to the surroundings and excludes all structures with the main spine smaller than $\simeq2\arcmin$, which are automatically filtered-out from the catalogue \citep{Schisano19}. The final catalogue contains more than 30000 filaments extracted from the whole Galactic plane.

A strong correlation between star-forming clumps and filamentary structures has been already observed, at least in nearby regions \citep{Schneider13,Konives15}. In fact, only 4 out of our 29 clumps do not have an association with a Hi-GAL filament in the \citet{Schisano19} catalogue. Three of these missing counterparts (20.717-0.01, 24.552+0.096, 25.254-0.166) are associated with IRDCs smaller than $2\arcmin$, while the IRDC 32.006-0.51 is fragmented in the Hi-GAL column density image and this is likely the reason why it was not identified in the \citet{Schisano19} catalogue. This ratio is in agreement with the findings of \citet{Schisano19}, that combined the whole Hi-GAL catalogue of filaments with the catalogue of $\simeq15000$ IRDCs from \citet{Peretto09} and found that only $\simeq28\%$ of the IRDCs do not have a counterpart in the Hi-GAL filaments, and in most cases because of the filtering of features smaller than $\simeq2\arcmin$.


We used the filament mask to extract the \co13\ $(1-0)$ data for each cloud from the GRS survey \citep[][]{Jackson06}. This survey observed a portion of the inner Galactic plane $(15\deg\leq l\leq55\deg, -1\deg\leq b\leq 1\deg)$ in the \co13 $(1-0)$ emission with a spatial resolution of $46\arcsec$ and a spectral resolution of $0.212$ km s$^{-1}$. The sensitivity is $\sigma_{T_{mb}}=0.26$ K, considering the main beam efficiency of 0.48 \citep{Roman-Duval10}.

From the GRS counterparts we excluded the filament associated with the source 28.905-0.534. The central velocity of the main CO emission is shifted $\simeq30$ km s$^{-1}$ with respect to the central velocity of the clump determined from the \n2h\ $(1-0)$ emission, and the CO emission around the clump velocity position is below the 4$\sigma$ level used as a lower limit for the molecular emission as discussed in Section \ref{sec:results}.

\begin{figure*}
\centering
\includegraphics[width=9.5cm]{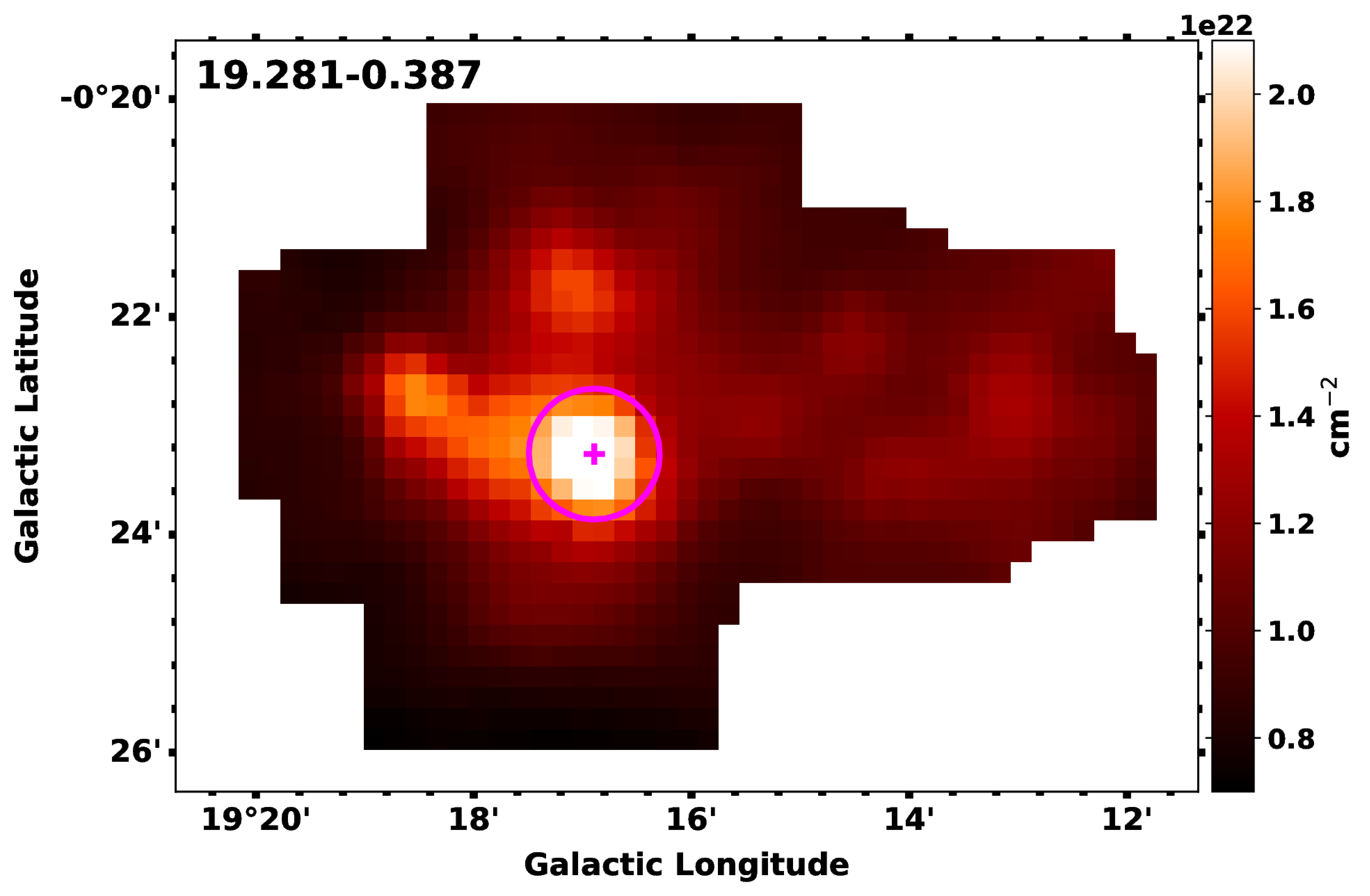} \qquad 
\includegraphics[width=7.2cm]{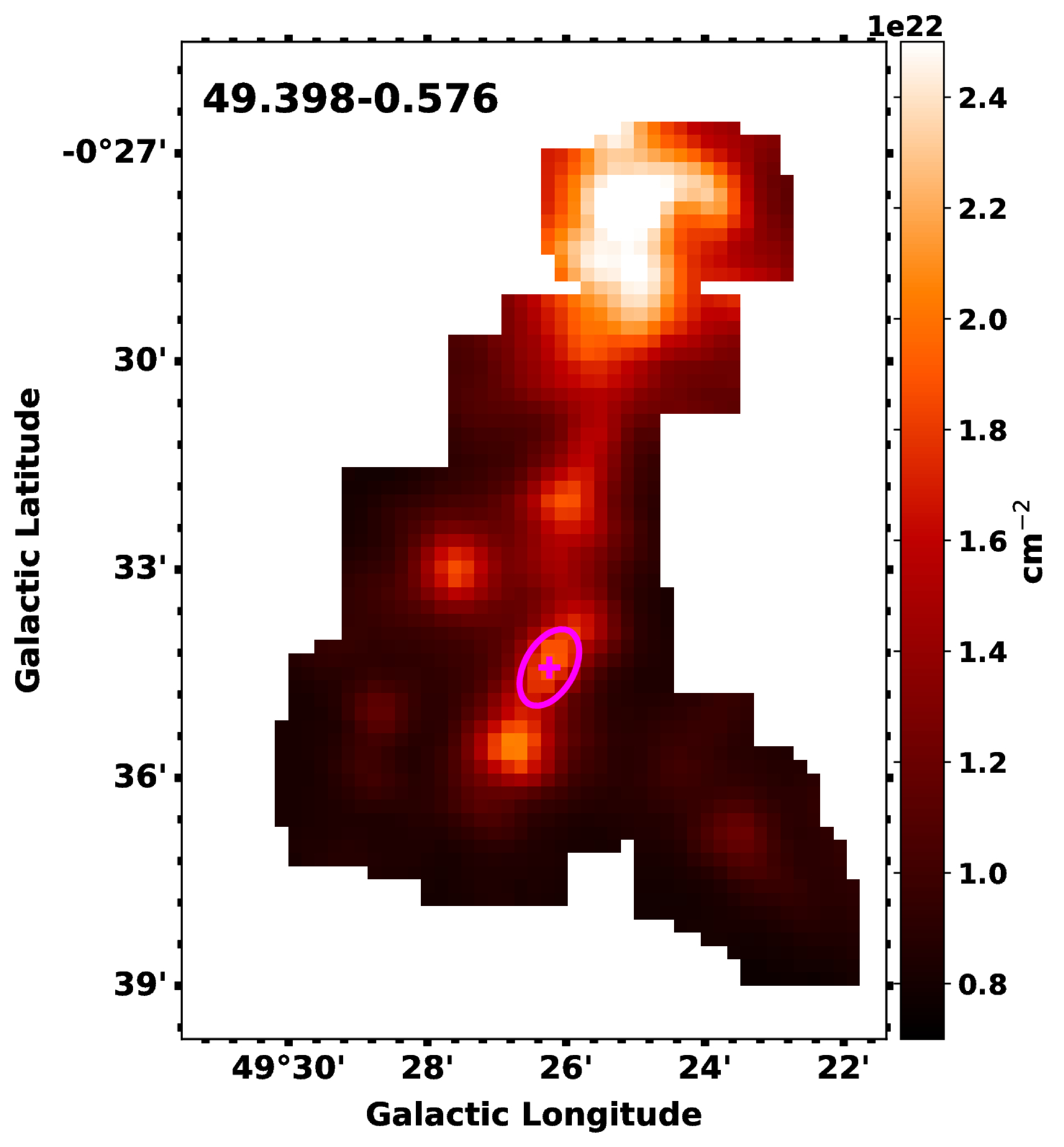}
\caption{Two examples of filaments extracted from the Hi-GAL column density maps and presented in the \citep{Schisano19} catalogue, associated with the surce 19.281-0.387 (left image) and 49.398-0.576 (right image) respectively. The magenta crosses and ellipses are in correspondence of the 70\mum\ quiet clumps associated with these two filaments.}    
\label{fig:fil_filament_clump}
\end{figure*}

The sample of filaments and clumps contains the physical parameters and the kinematics of 24 out of the 29 sources. Two examples of the filaments extracted from the Hi-GAL column density maps and the embedded clumps for which we have kinematic information are in Figure \ref{fig:fil_filament_clump}.

To estimate the properties of the filaments such as optical depth and mass from the \co13\ data, we need to evaluate the excitation temperature \tex. The excitation temperature is evaluated under the local thermal equilibrium assumption combining the \12co\ and \co13\ data taken from the FUGIN survey \citep{Umemoto17}. This survey has been carried out with the Nobeyama 45m telescope and observed the \12co, \co13\ and C$^{18}$O $(1-0)$ transitions simultaneously across 156 deg$^{2}$ of the inner Galactic Plane in a 2 degrees wide longitude strip around $b=0\deg$. Our clumps and filaments are all in the I quadrant, where the FUGIN survey covers the longitudes $10\deg\leq l\leq50\deg$. The noise is 0.24 K per channel and 0.12 K per channel for the \12co\  and the \co13\ cubes respectively. The FUGIN survey has a better spatial resolution than the GRS survey, $\simeq20\arcsec$, however it has a lower spectral resolution, 0.65 km s$^{-1}$. 

We therefore used both the \12co\ and \co13\ FUGIN cubes to evaluate the excitation temperature as discussed in Section \ref{sec:results}, but we take advantage of the finer spectral resolution of the GRS data to estimate the physical parameters and the velocity dispersion of the clouds.

\section{Results}\label{sec:results}
The goal of this work is to investigate whether the dynamics of the massive and dense clumps that can lead to the formation of high-mass stars are peculiar or not, compared to the lower density counterparts. The surface density threshold discussed in the literature above which a clump will likely form high-mass objects varies from $\Sigma_{thres}=0.05$ g cm$^{-2}$ \citep{Urquhart14} to $\Sigma_{thres}=0.1$ g cm$^{-2}$ \citep{Tan14,Traficante17_PII}, and up to $\Sigma_{thres}=1$ g cm$^{-2}$ if we consider the radiative feedback needed to prevent the fragmentation and the formation of the massive objects in some models \citep{Krumholz08,Federrath17,Guszejnov18}. Based on these numbers, we divide our sample in three different groups:

\begin{itemize}
\item[$\Sigma_{low}$:] clumps with a (relatively) low surface density, below the minimum threshold assumed for high-mass star-formation, $\Sigma\leq0.05$ g cm$^{-2}$; 

\item[$\Sigma_{int}$:] clumps with higher surface density than the ones in the $\Sigma_{l}$ group, but still below the 0.1 g cm$^{2}$ critical value. These are defined intermediate-density clumps; 

\item[$\Sigma_{high}$:] these are clumps with the surface density above the surface density threshold of $\Sigma=0.1$ g cm$^{-2}$.  


\end{itemize}

We aim to investigate any possible difference between regions in these surface density regimes, in particular between clumps and filaments in the $\Sigma_{low}$ and $\Sigma_{high}$ groups, with a somehow intermediate behavior for the regions in the $\Sigma_{int}$ group.

\subsection{Clumps analysis}
The physical properties of each of the new 13 clumps have been re-derived with the same approach used for the massive 70\mum\ quiet clumps in \citet{Traficante17_PI} for consistency. Therefore, we performed a dedicated run to extract the fluxes at all Hi-GAL wavelengths with the \Hyp\ routine \citep{Traficante15b} including also the emission at 870\mum\ from ATLASGAL and at 1100\mum\ from the BGPS. \Hyp\ performs a 2d-Gaussian fit at 250\mum\ of each source that it is used to determine the two semi-axes (the two FWHMs of the fit) and the position angle of the ellipse. These parameters define the radius $R$ of the clump and the region of integration of the source fluxes at all wavelengths.

The physical parameters of each clump are determined with a single-temperature grey-body fit of the spectral energy distribution using all wavelengths in the range $160\leq\lambda\leq1100$\mum. If the observations in the ATLASGAL and BGPS surveys are missing, we perform the SED fitting only on the Hi-GAL wavelengths ($160\leq\lambda\leq500$\mum). This is the case for six sources. The greybody fit is done assuming a fixed spectral index $\beta=2.0$ and with the uncertainties associated to each flux equal to 20$\%$ of the integrated fluxes in the 70-500\mum\ wavelength range and to 40$\%$ of the integrated fluxes for the 870 and 1100\mum\ values, to include also the uncertainties due to the combination of different surveys with the Hi-GAL fluxes \citep{Traficante17_PI}. The masses of these clumps are in the range $25\leq M\leq531$ M\sun. We obtain surface densities $\Sigma=M/(\pi R^{2})$ in the range $0.006\leq\Sigma\leq0.07$ g cm$^{-2}$. The uncertainties associated with each measurements are derived as in \citet{Traficante18}, that made an extensive analysis to evaluate the source of uncertainties for each parameter.  A summary of the source parameters for all the 29 clumps used in this work are in Table \ref{tab:clump_params}. The clumps have been divided in three groups based on the value of their surface density. We obtain 10, 8 and 11 clumps belonging, respectively, to the $\Sigma_{low}$, $\Sigma_{int}$ and $\Sigma_{high}$ group.

In order to study the dynamics of each clump, we need to estimate their velocity dispersion. In agreement with the analysis done for the 16 clumps in \citet{Traficante17_PI}, we first determined the \n2h\ $(1-0)$ spectrum averaged in all pixels within the ellipse defined by the 2d-Gaussian fit of each source, then we estimated the velocity dispersion from a hyperfine fitting of the average spectrum performed with standard CLASS routines. As showed in Table \ref{tab:clump_params}, all these clumps are supersonic, with $\sigma\geq0.39$ km s$^{-1}$ (the H$_{2}$ thermal velocity at temperature T=10 K is $\simeq0.2$ km s$^{-1}$), including two clumps that are mildly supersonic (25.254-0.166, $\sigma=0.29$ km s$^{-1}$, and 49.398-0.576, $\sigma=0.34$ km s$^{-1}$). There are no evident differences between the three groups, despite the difference in mass between clumps can be up to 1 order of magnitude.

\begin{center}
\begin{table*}
\centering
\begin{tabular}{c|c|c|c|c|c|c|c|c|c|c}
\hline
\hline
 Clump & $M_{clump}$ & $R_{cl}$ & $\Sigma_{cl}$ & $\sigma_{clump}$ & $\alpha_{vir}$ & distance & RA & Dec & Group \\
  & (M\sun) & (pc) & (g cm$^{-2}$) & (km s$^{-1}$) & & (kpc) & ($\deg$) & ($\deg$) & \\
\hline

28.905-0.534  &      25(  7)  &    0.54(0.14)  &   0.006(0.002)  &    0.45(0.13)  &    4.85(3.15)  & 4.34  & 18:46:04.3  & -3:47:24  &  $\Sigma_{low}$ \\
       24.552+0.096  &      47( 13)  &    0.55(0.14)  &   0.011(0.004)  &    0.64(0.19)  &    5.42(3.52)  & 3.44  & 18:35:42.2  & -7:24:21  &  $\Sigma_{low}$ \\
       25.254-0.166  &      59( 16)  &    0.59(0.15)  &   0.011(0.004)  &    0.29(0.09)  &    0.94(0.61)  & 3.89  & 18:37:57.6  & -6:54:14  &  $\Sigma_{low}$ \\
       49.732-0.471  &     110( 30)  &    0.79(0.20)  &   0.012(0.004)  &    0.43(0.13)  &    1.51(0.98)  & 5.43  & 19:24:33.1  & 14:41:59  &  $\Sigma_{low}$ \\
        20.717-0.01  &     102( 30)  &    0.50(0.12)  &   0.027(0.010)  &    0.97(0.29)  &    5.33(3.46)  & 4.05  & 18:28:57.6  &-10:51:32  &  $\Sigma_{low}$ \\
        28.19-0.192  &     219( 63)  &    0.71(0.18)  &   0.029(0.010)  &    0.59(0.18)  &    1.30(0.85)  & 4.47  & 18:43:29.0  & -4:19:15  &  $\Sigma_{low}$ \\
       53.784-0.138  &     102( 35)  &    0.48(0.12)  &   0.030(0.010)  &    0.54(0.16)  &    1.57(1.02)  & 4.96  & 19:31:27.1  & 18:25:40  &  $\Sigma_{low}$ \\
       45.531+0.042  &     212( 62)  &    0.67(0.17)  &   0.031(0.011)  &    0.55(0.16)  &    1.10(0.71)  & 4.51  & 19:14:33.1  & 11:12:50  &  $\Sigma_{low}$ \\
       23.076-0.209  &     159( 44)  &    0.55(0.14)  &   0.035(0.012)  &    0.58(0.17)  &    1.34(0.87)  & 3.72  & 18:34:04.1  & -8:51:25  &  $\Sigma_{low}$ \\
       26.432-0.662  &     221( 86)  &    0.64(0.16)  &   0.036(0.013)  &    0.81(0.24)  &    2.20(1.43)  & 3.92  & 18:41:40.3  & -6:04:55  &  $\Sigma_{low}$ \\
       35.608+0.111  &     176( 49)  &    0.50(0.12)  &   0.047(0.016)  &    0.53(0.16)  &    0.92(0.60)  & 3.16  & 18:55:55.7  &  2:26:02  &  $\Sigma_{low}$ \\

\\

       30.357-0.837$^{1}$  &     371(110)  &    0.67(0.17)  &   0.055(0.019)  &    0.57(0.17)  &    0.68(0.44)  
& 4.30  & 18:49:40.6  & -2:39:46  &  $\Sigma_{int}$ \\
       53.361+0.042  &      36(  9)  &    0.21(0.05)  &   0.055(0.019)  &    0.39(0.12)  &    0.99(0.65)  & 2.04  & 19:29:48.2  & 18:08:02  &  $\Sigma_{int}$ \\
       30.131-0.644  &     379(101)  &    0.66(0.17)  &   0.058(0.020)  &    0.56(0.17)  &    0.63(0.41)  & 4.59  & 18:48:38.2  & -2:47:06  &  $\Sigma_{int}$ \\
        32.006-0.51$^{1}$  &     448(127)  &    0.70(0.17)  &   0.061(0.021)  &    0.31(0.09)  &    0.18(0.12)  & 4.24  & 18:51:34.1  & -1:03:25  &  $\Sigma_{int}$ \\
       15.631-0.377$^{1}$  &     268( 78)  &    0.54(0.14)  &   0.061(0.021)  &    0.30(0.09)  &    0.21(0.14)  & 3.47  & 18:20:29.0  &-15:31:26  &  $\Sigma_{int}$ \\
       49.398-0.576  &     531(176)  &    0.74(0.18)  &   0.065(0.023)  &    0.34(0.10)  &    0.18(0.12)  & 5.47  & 19:24:18.5  & 14:22:30  &  $\Sigma_{int}$ \\
       28.792+0.141$^{1}$  &     449(128)  &    0.61(0.15)  &   0.080(0.028)  &    0.99(0.30)  &    1.55(1.01)  & 4.62  & 18:43:08.9  & -3:36:17  &  $\Sigma_{int}$ \\
        49.433-0.22  &     385(108)  &    0.53(0.13)  &   0.091(0.032)  &    0.82(0.25)  &    1.07(0.70)  & 4.32  & 19:22:52.6  & 14:30:43  &  $\Sigma_{int}$ \\
       25.982-0.056$^{1}$  &     888(274)  &    0.80(0.20)  &   0.092(0.032)  &    0.69(0.21)  &    0.50(0.32)  & 5.00  & 18:38:54.5  & -6:12:32  &  $\Sigma_{int}$ \\
       33.332-0.531  &    1266(350)  &    0.94(0.23)  &   0.095(0.033)  &    0.86(0.26)  &    0.64(0.41)  & 5.37  & 18:54:03.8  &  0:07:01  &  $\Sigma_{int}$ \\
       
\\
       
       19.281-0.387$^{1}$  &     700(206)  &    0.67(0.17)  &   0.104(0.036)  &    0.47(0.14)  &    0.25(0.16)  & 3.82  & 18:27:33.8  &-12:18:17  & $\Sigma_{high}$ \\
       34.131+0.075$^{1}$  &     480(141)  &    0.55(0.14)  &   0.106(0.037)  &    0.74(0.22)  &    0.72(0.47)  & 3.56  & 18:53:21.6  &  1:06:14  & $\Sigma_{high}$ \\
       23.271-0.263$^{1}$  &     997(296)  &    0.72(0.18)  &   0.128(0.045)  &    0.94(0.28)  &    0.75(0.49)  & 5.21  & 18:34:37.9  & -8:40:44  & $\Sigma_{high}$ \\
       31.946+0.076$^{1}$  &    1431(439)  &    0.82(0.20)  &   0.142(0.050)  &    1.19(0.36)  &    0.94(0.61)  & 5.51  & 18:49:22.3  &  0:**:31  & $\Sigma_{high}$ \\
       22.756-0.284$^{1}$  &     655(193)  &    0.55(0.14)  &   0.144(0.050)  &    0.95(0.28)  &    0.88(0.57)  & 4.43  & 18:33:49.0  & -9:13:04  & $\Sigma_{high}$ \\
        22.53-0.192$^{1}$  &    1579(487)  &    0.80(0.20)  &   0.164(0.057)  &    1.25(0.37)  &    0.92(0.60)  & 5.77  & 18:32:59.8  & -9:20:02  & $\Sigma_{high}$ \\
       28.537-0.277$^{1}$  &    1152(326)  &    0.67(0.17)  &   0.171(0.060)  &    0.78(0.23)  &    0.41(0.27)  & 4.96  & 18:44:22.1  & -4:01:40  & $\Sigma_{high}$ \\
       28.178-0.091$^{1}$  &    2092(610)  &    0.85(0.21)  &   0.193(0.067)  &    1.07(0.32)  &    0.54(0.35)  & 5.35  & 18:43:02.6  & -4:14:52  & $\Sigma_{high}$ \\
       24.013+0.488$^{1}$  &    1957(559)  &    0.81(0.20)  &   0.198(0.069)  &    0.91(0.27)  &    0.40(0.26)  & 5.18  & 18:33:18.5  & -7:42:25  & $\Sigma_{high}$ \\
       25.609+0.228$^{1}$  &    3098(953)  &    0.97(0.24)  &   0.219(0.077)  &    1.05(0.31)  &    0.40(0.26)  & 5.57  & 18:37:10.6  & -6:23:31  & $\Sigma_{high}$ \\
       18.787-0.286$^{1}$  &    1915(549)  &    0.69(0.17)  &   0.267(0.094)  &    1.07(0.32)  &    0.48(0.31)  & 4.36  & 18:26:15.4  &-12:41:34  & $\Sigma_{high}$ \\
\hline
\end{tabular}

\begin{tablenotes}
\scriptsize
\item $^{1}$ From \citet{Traficante17_PI,Traficante17_PII}
\end{tablenotes}

\caption{Physical and kinematic parameters of the 29 clumps used in this work, combining the 13 clumps observed with IRAM 30m under the project 133-15 and the 16 objects presented in \citet{Traficante17_PI,Traficante17_PII}. The sources are divided in three groups, as described in Section \ref{sec:results}, and ordered by increasing values of their surface density within each group. Col. 1: Clump name; Cols. 2-4: clump mass, radius and surface density derived from the \textit{Hyper} fitting at 250\mum; Col. 5: velocity dispersion measured from the hyperfine fitting of the \n2h\ $(1-0)$ spectrum; Col. 6: virial parameter; Col. 7: distance of the source obtained from the \citet{Traficante15a} catalogue; Cols. 8-9: equatorial coordinates of the 250\mum\ centroid obtained from the Gaussian fitting; Col. 10: group based on the clump surface density as discussed in Section \ref{sec:results}.}
\label{tab:clump_params}
\end{table*}
\end{center}

\subsubsection{The dynamics of star-forming clumps}\label{sec:clumps_dynamics}
In this Section we study the dynamics of our sample of clumps and we divide and compare the results between the three different groups. 

We start with a useful parameter to investigate the energy budget of a clump, the virial parameter \avir, which describes the ratio between the gravitational and the kinetic energy of a defined region. Here, we evaluate \avir\ following the \citet{Bertoldi92} definition of $\alpha_{vir}=2aE_{k}/E_{G}=a5\sigma^2 R/(GM)$, where $E_{k}$ and $E_{G}$ are the kinetic and gravitational energy respectively, $M$ and $R$ are the mass and radius of the clump respectively, G is the gravitational constant and $a$ is a constant which includes modifications due to non-spherical and inhomogeneous density distributions. We assume for simplicity $a=1$, although accounting for a more realistic mass and kinematic distribution of the gas in the clumps and their surroundings can lead to different values of $a$ and strongly affect the values of $E_{k}$ and $E_{G}$ \citep{Federrath12,Beaumont13}.

\begin{figure}
\centering
\includegraphics[width=8.2cm]{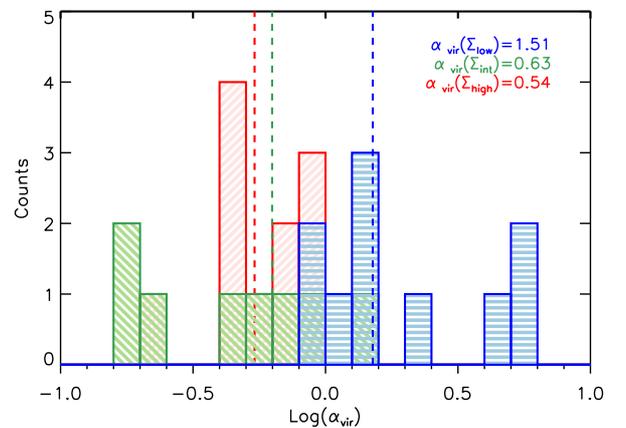}
\caption{Histogram of the virial parameters of the 70\mum\ quiet clumps divided for the three groups as discussed in Section \ref{sec:results}.}    
\label{fig:histo_alpha_vir}
\end{figure}

In Figure \ref{fig:histo_alpha_vir} we show the histogram of the virial parameter divided in the three groups. Blue bins are used for the $\Sigma_{low}$ sources, green bins for the $\Sigma_{int}$ sources and red bins for the $\Sigma_{high}$ sources. The dotted lines are the median values for each group, which are \avir=[1.51, 0.63, 0.54] for the [$\Sigma_{low}$, $\Sigma_{int}$, $\Sigma_{high}$] clumps respectively. The low-density clumps have a wider distribution of the virial parameter, with the large majority of them with values \avir$>1$ and with few objects (4 out of 10) that have \avir$>2$. A value of the virial parameter larger than 2 suggests that these regions are gravitationally unbound \citep{Kauffmann13,Tan14} and the gravity alone cannot overcome the strong non-thermal motions fed by the turbulence in the interstellar medium, although this value may be over-estimated as a result of an underestimation of the gravitational energy budget of a given region \citep{Ballesteros-Paredes18}. On the other hand, super-critical values of the virial parameter, i.e. \avir$<<1$ are considered as evidence that the clumps are significantly gravitationally bound and therefore prone to collapse \citep{Urquhart18}, but also in this case the value of \avir\ obtained from the observations of apparent super-virial clumps may be misleading due to a bias in the estimation of the local non-thermal motions \citep{Traficante18_PIII}. These regions should indeed be naturally close to the virial value \citep{Lee16a}, and often appear virialized \citep{Ballesteros-Paredes11}.

\begin{figure}
\centering
\includegraphics[width=8.5cm]{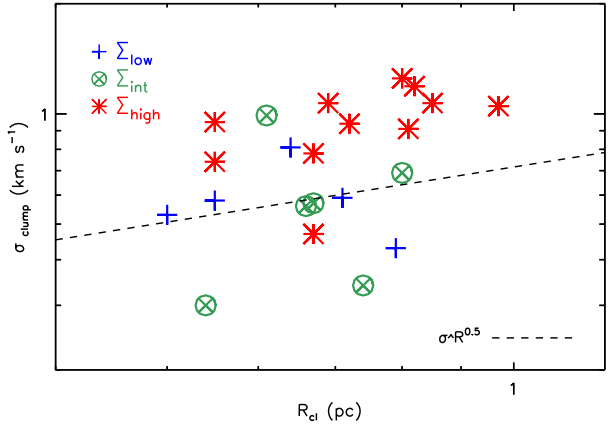}
\caption{Larson's relation for the clumps divided in the three groups. The black-dashed line is the classical Larson's relation $\sigma\propto R^{\delta}$ with $\delta=0.5$ and normalized to 0.72 km s$^{-1}$ at 1 pc \citep{Heyer15}.}    
\label{fig:Larson_first_clumps}
\end{figure}

In general, the virial parameter gives a rough estimate of the energy balance of each region, and it is not possible to distinguish if the observed non-thermal motions are dominated by local turbulence or by the gravitational collapse from solely its value. In the former scenario, we expect to observe a velocity dispersion-size relation, $\sigma\propto R^{\delta}$, that resembles the turbulent cascade of energy when $\delta=0.5$ \citep{Larson81,Heyer04,McKee07,Federrath13}. In \citet{Traficante18} we have already discussed how this relation breaks down at the clump scales. We draw similar conclusions from our sample of objects, as shown in Figure \ref{fig:Larson_first_clumps}, although our sample covers a limited range of radii. A correlation is possibly present in the high density clumps (Pearson's coefficient $\rho=0.53$), but it is mild in the $\Sigma_{int}$ clumps ($\rho=0.29$) and absent in the $\Sigma_{low}$ ones ($\rho=-0.22$). The combined sample of high and intermediate density clumps, $\Sigma_{int+high}$, that together form a sample of 18 objects, also have a mild correlation ($\rho=0.55$).

If it is gravity that is playing a central role in driving the observed non-thermal motions, then the clumps should follow the generalized form of the Larson's relation, $\sigma^{2}/r\propto\Sigma$ relation \citep{Heyer09,Ballesteros-Paredes11,Traficante17_PII}. Massive clumps seem to follow this generalized form \citep{Camacho16}, which is interpreted as evidence of a hierarchical, global collapse driven by the self-gravity \citep{Ballesteros-Paredes11,Vazquez-Semadeni19}.

\begin{figure}
\centering
\includegraphics[width=8.8cm]{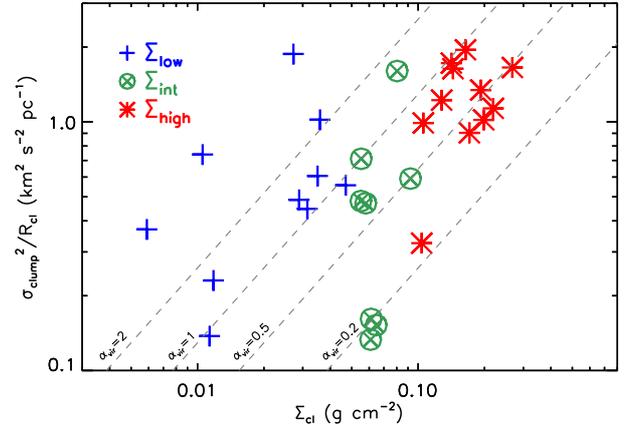}
\caption{The quantity $\sigma^{2}/R$ as a function of the surface density $\Sigma$ for the clumps belonging to the three groups. The relation changes considering only the clumps belonging to the $\Sigma_{high}$ and $\Sigma_{int}$ groups, or including also the clumps in the $\Sigma_{low}$ group. The grey-dotted lines are in correspondence of constant values of the virial parameter.} 
\label{fig:ak_ag_all}
\end{figure}

In Figure \ref{fig:ak_ag_all} we show the $\sigma^{2}/r\propto\Sigma$ relation for our clumps. If we focus on the clumps belonging to the intermediate and high surface density groups, they show a good degree of correlation, with Pearson's coefficient $\rho_{\Sigma_{int+high}}=0.65$. The relation changes as we introduce the low-density clumps. These clumps are much sparser in the parameter space, as a consequence of the large spread in the values of their virial parameter. The Pearson's coefficient of the $\Sigma_{low}$ clumps alone is also relatively low, $\rho=0.31$, suggesting that the behavior of these objects is different from the rest of the sample.

A similar way to explore if gravity plays a dominant role in the collapse is through the velocity dispersion-surface density diagram. This relation is similar to the generalized Larson relation but it has a great advantage, both $\Sigma$ and $\sigma$ are two distance-independent quantities, so they are not affected by one of the strongest sources of uncertainty in their estimation. If the non-thermal motions are driven by gravity, they are expected to correlate, keeping the region in a pseudo-virial state \citep{Ballesteros-Paredes11}. 

\begin{figure}
\centering
\includegraphics[width=8.8cm]{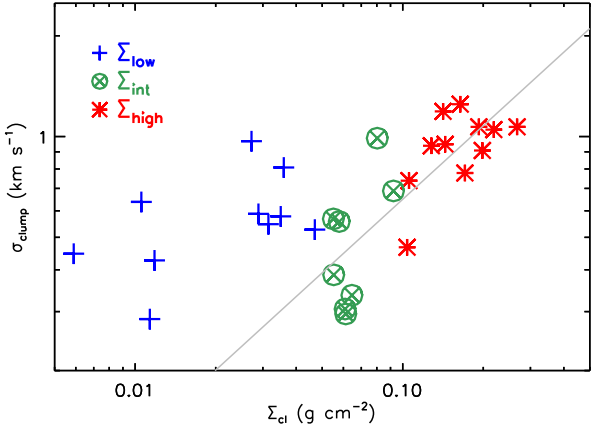}
\caption{Velocity dispersion of the clumps as a function of the surface density. The grey line is the linear fit to the log-log space of clumps belonging to the $\Sigma_{int}$ and $\Sigma_{high}$ groups. The grey-dotted horizontal line is the average value of the velocity dispersion for the clumps in the $\Sigma_{low}$ group.}    
\label{fig:Sigma_sigma_all}
\end{figure}

In Figure \ref{fig:Sigma_sigma_all} we report the $\Sigma$ vs. $\sigma$ diagram. The correlation is good for the clumps in the $\Sigma_{int+high}$ groups, with a Pearson's coefficient of 0.76. The solid line is the fit of the $\Sigma_{int+high}$ clumps done with the \texttt{linefit} IDL routine in the Log-Log space, which gives a scaling law $\sigma\propto\Sigma^{0.73}$. Analogously to what happened in the $\sigma^{2}/r\propto\Sigma$ plot, if we consider also the low-density clumps to the diagram the correlation breaks down. The dotted grey line in the plot is the average value of the non-thermal motions in the $\Sigma_{low}$ clumps. The line bumps into the result of the fit for the clumps of the $\Sigma_{int+high}$ groups at a surface density $\Sigma\simeq0.08$ g cm$^{-2}$. This value is very close to the critical value of $\Sigma\simeq0.1$  g cm$^{-2}$ above which all the 70\mum\ quiet clumps show evidence of dynamical activities \citep{Traficante17_PI}. In some of these $\Sigma_{high}$ regions the \hco\ $(1-0)$ and HNC $(1-0)$ spectra exhibit clear blue-asymmetries interpreted as evidence of global, parsec-scale collapse \citet{Fuller05,Rygl13,Kirk13,Traficante17_PI}, although even the red-asymmetries may be the result of the inflow of material \citep{Smith13}.

\begin{figure*}
\centering
\includegraphics[width=4.2cm]{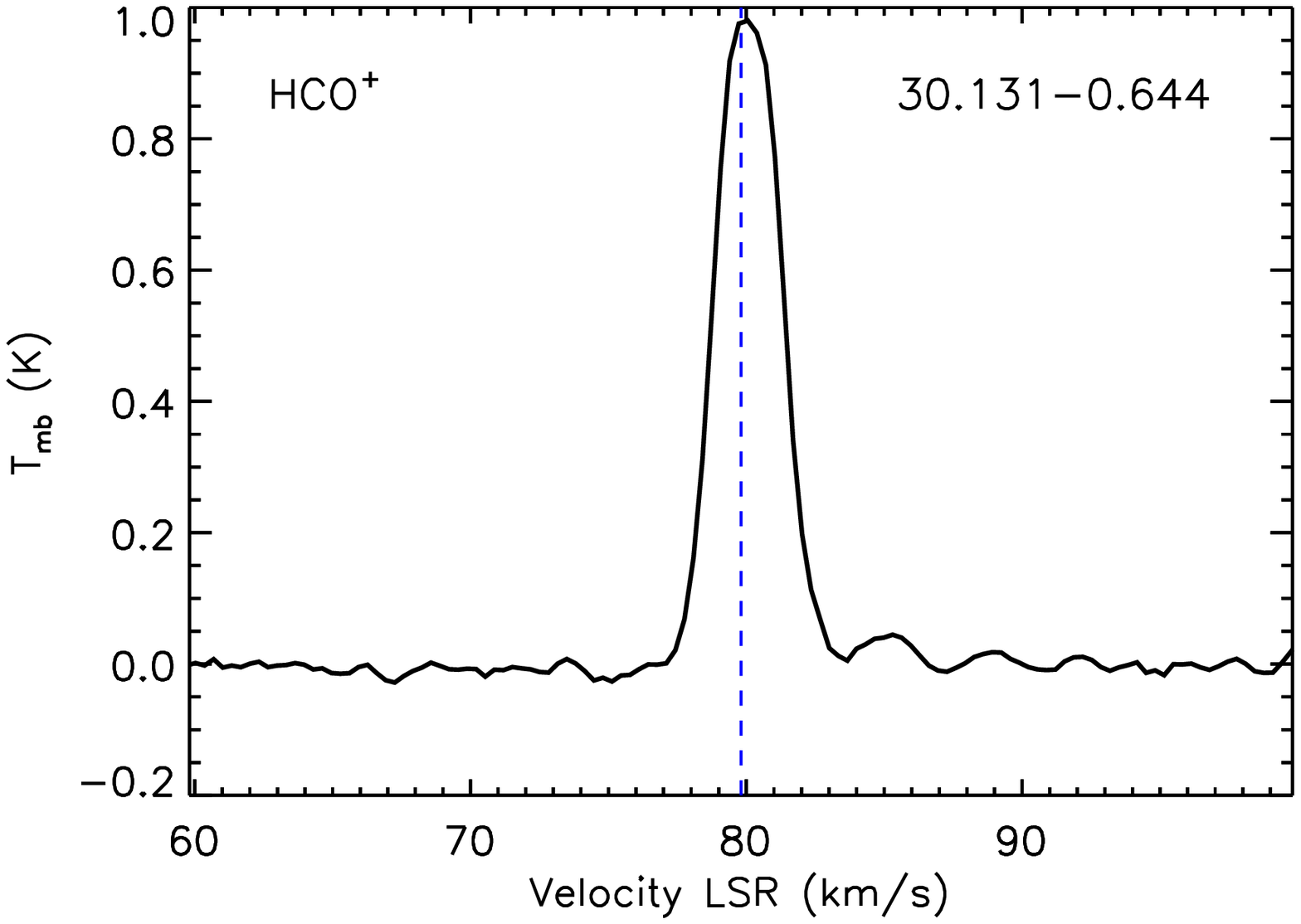}
\includegraphics[width=4.2cm]{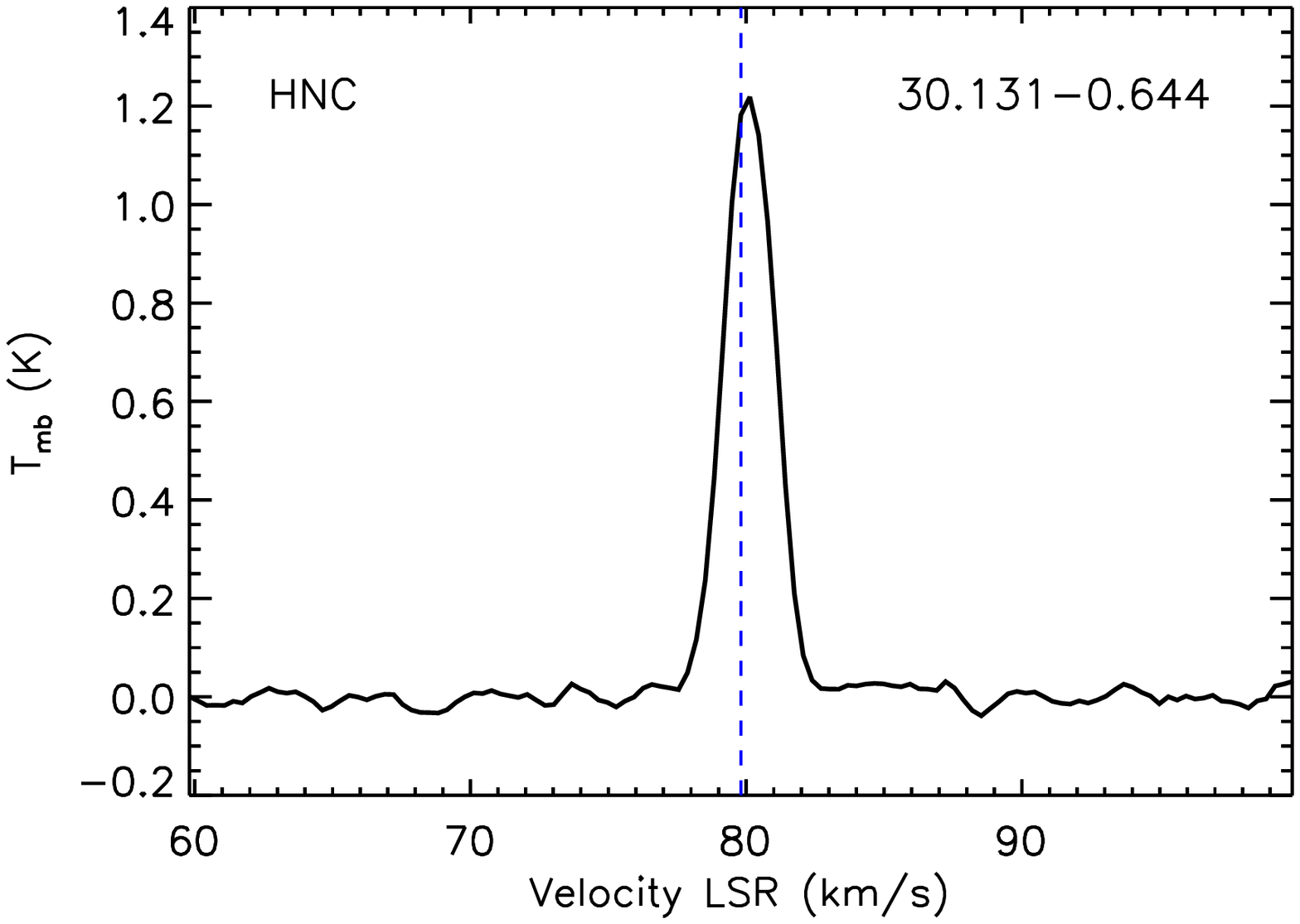}
\includegraphics[width=4.2cm]{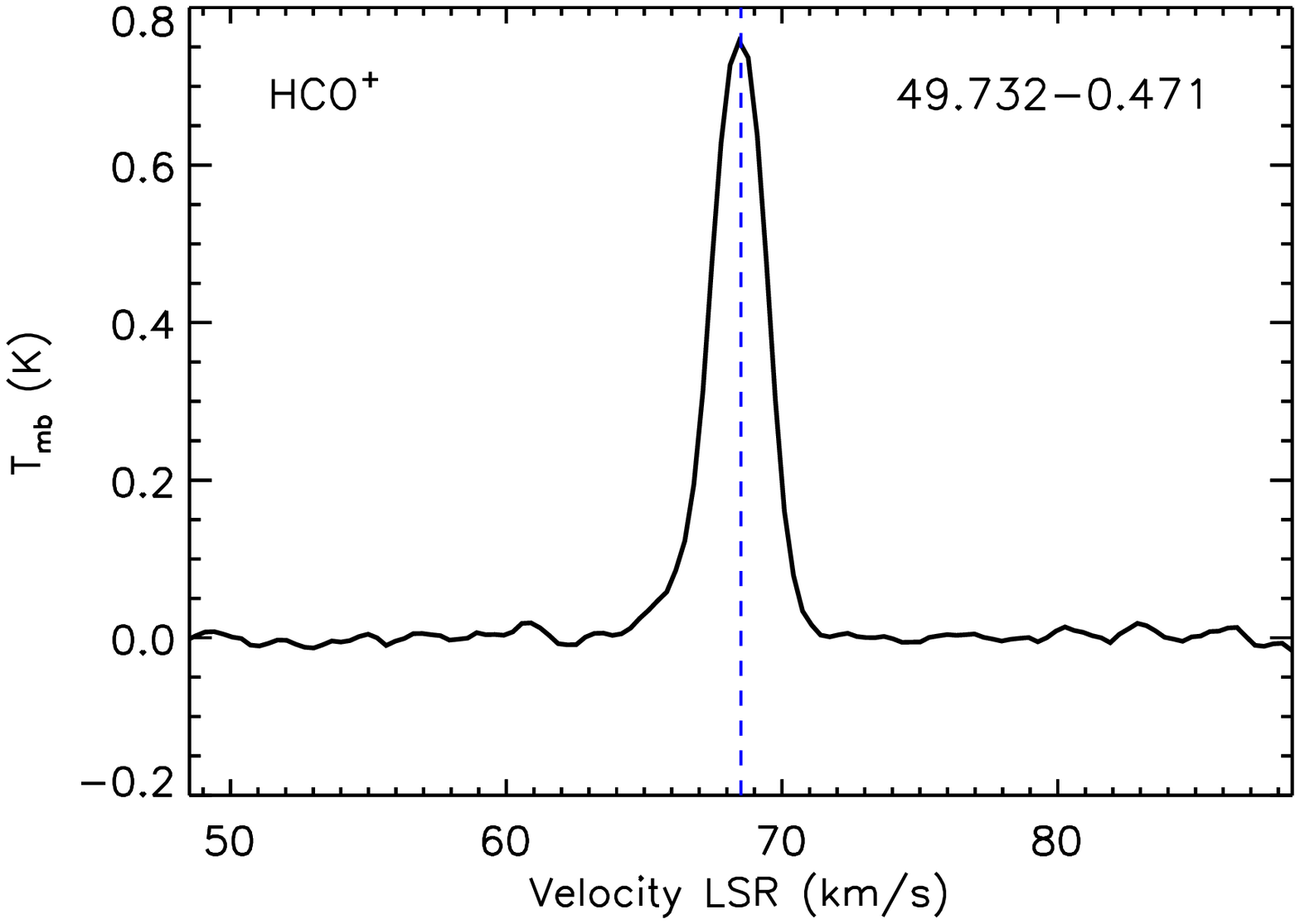}
\includegraphics[width=4.2cm]{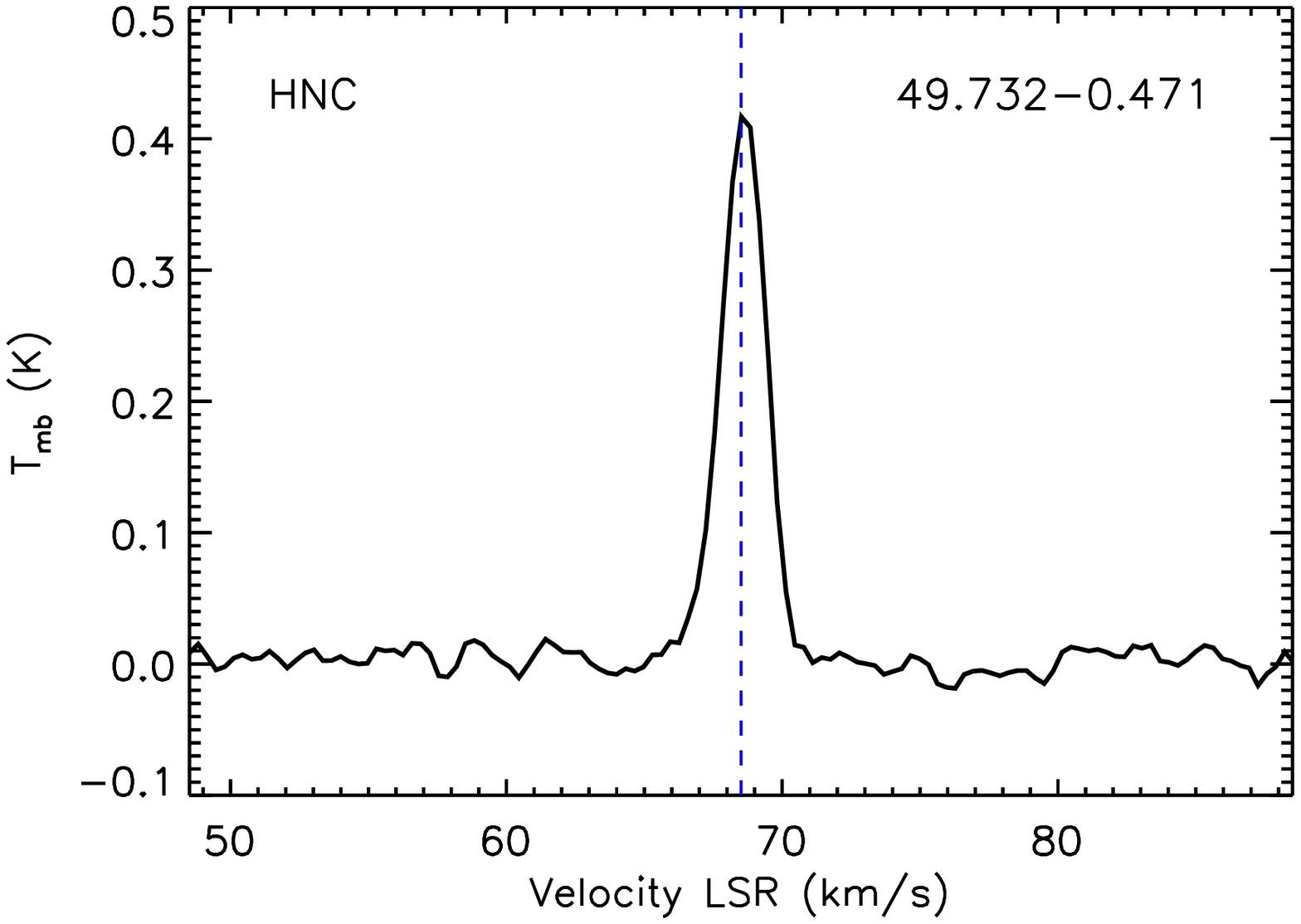}
\includegraphics[width=4.2cm]{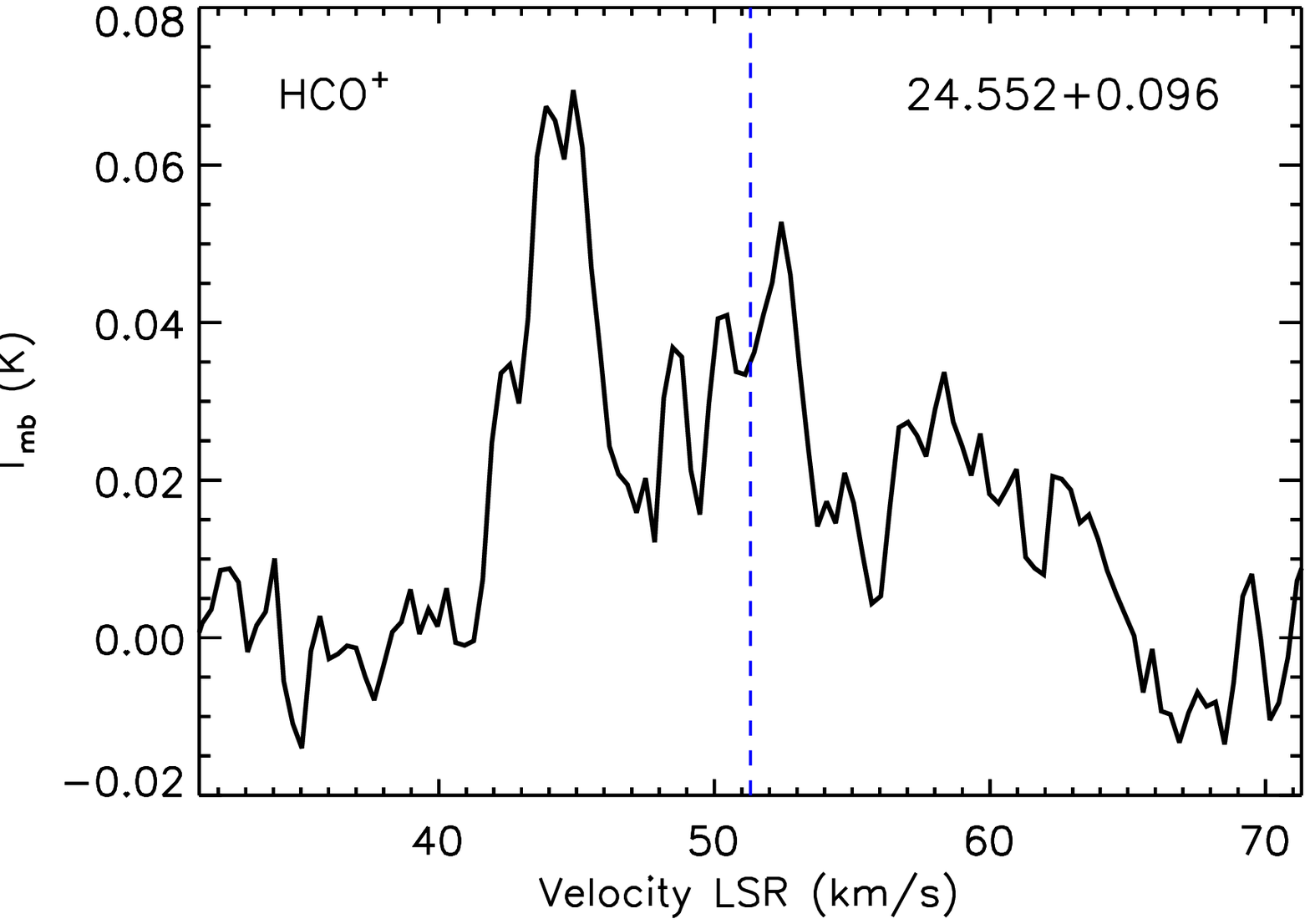}
\includegraphics[width=4.2cm]{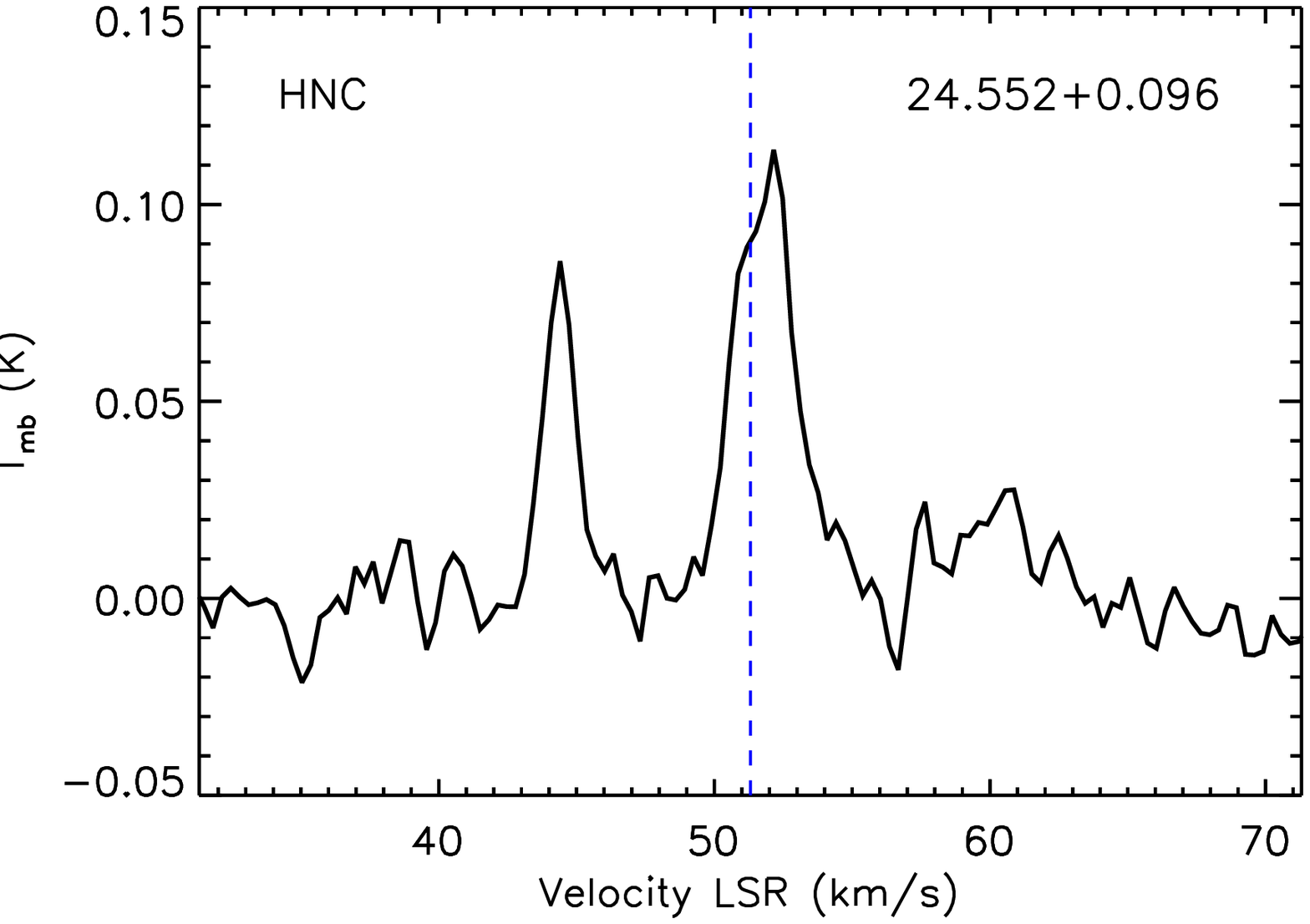}
\includegraphics[width=4.2cm]{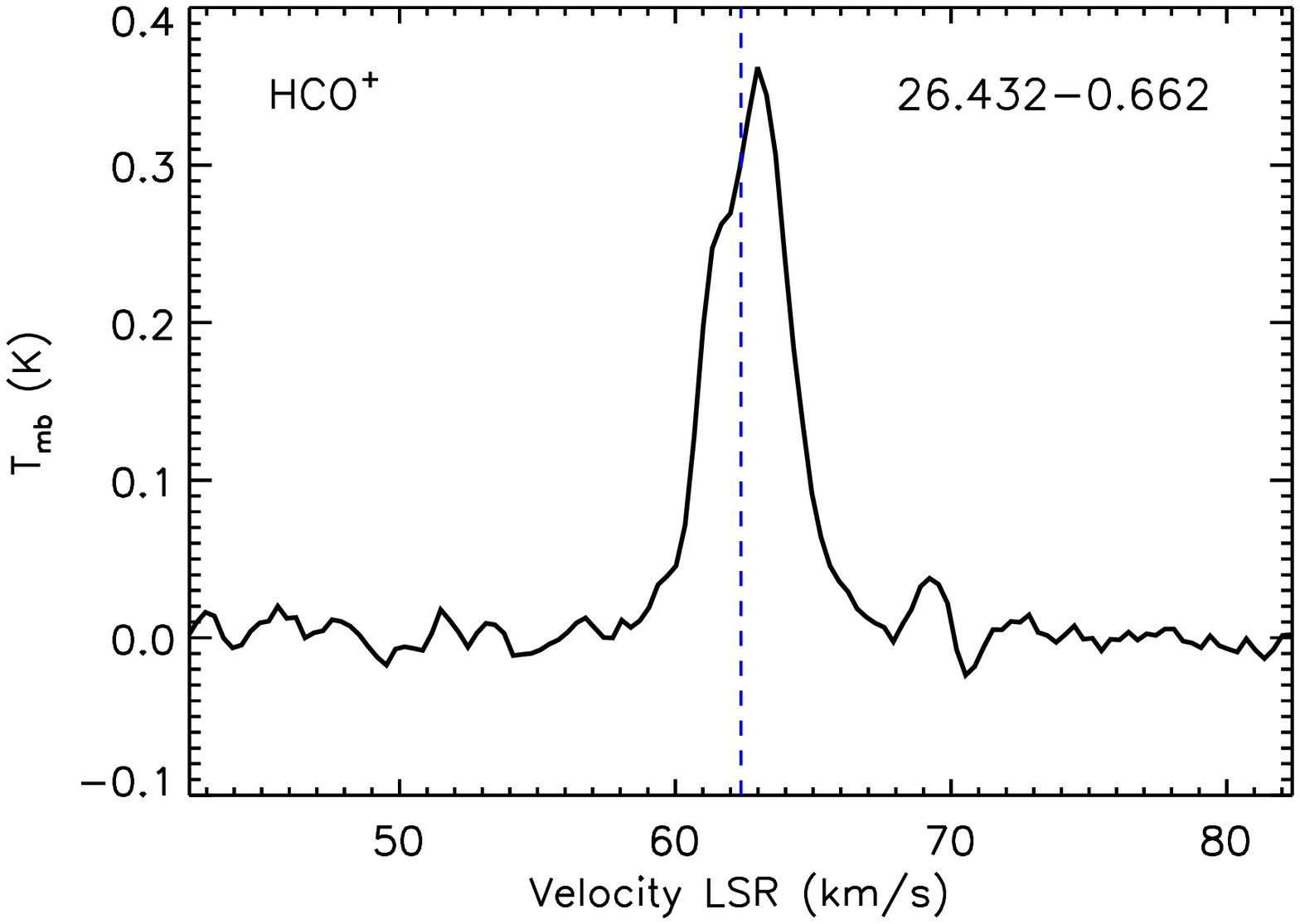}
\includegraphics[width=4.2cm]{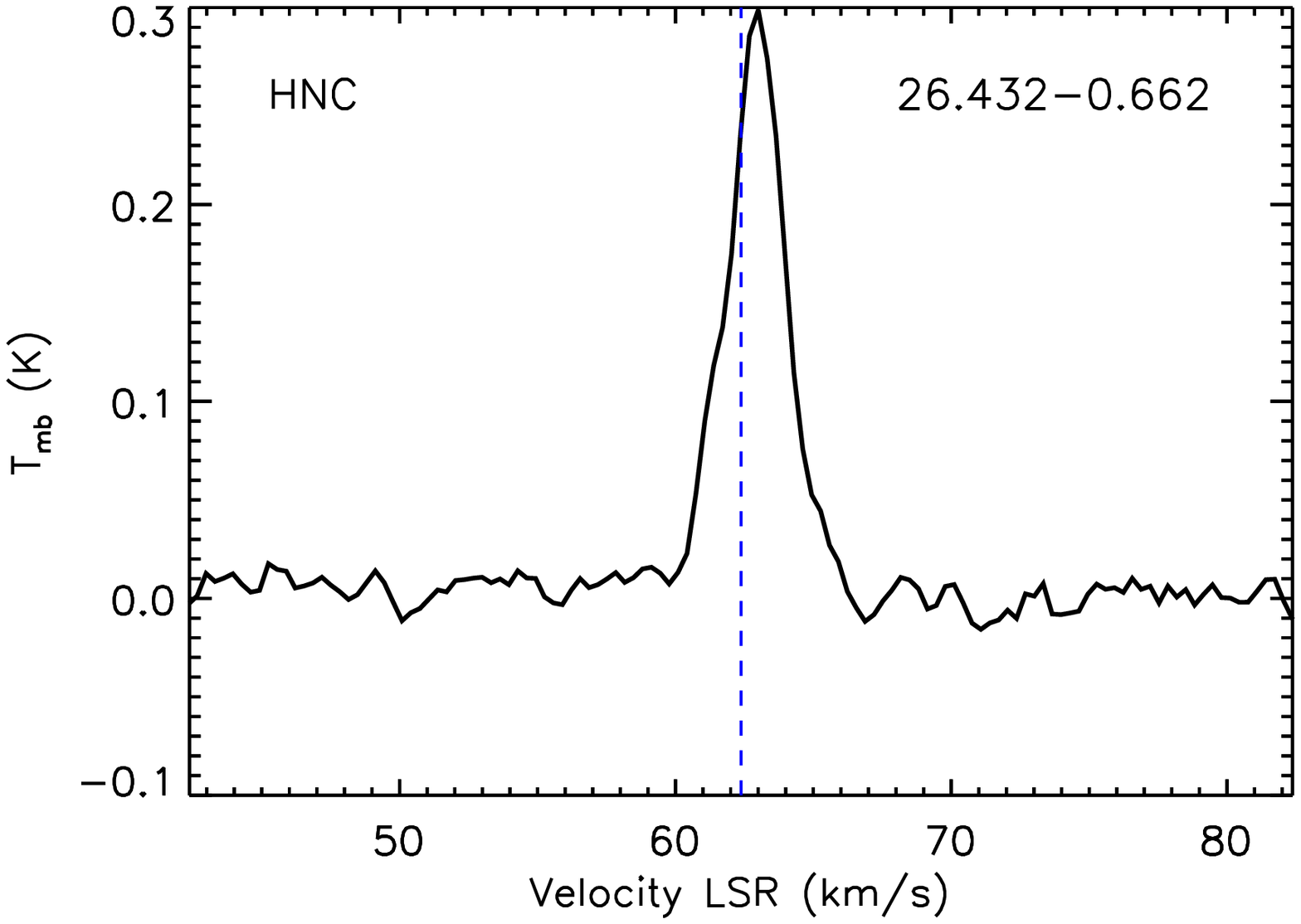}
\includegraphics[width=4.2cm]{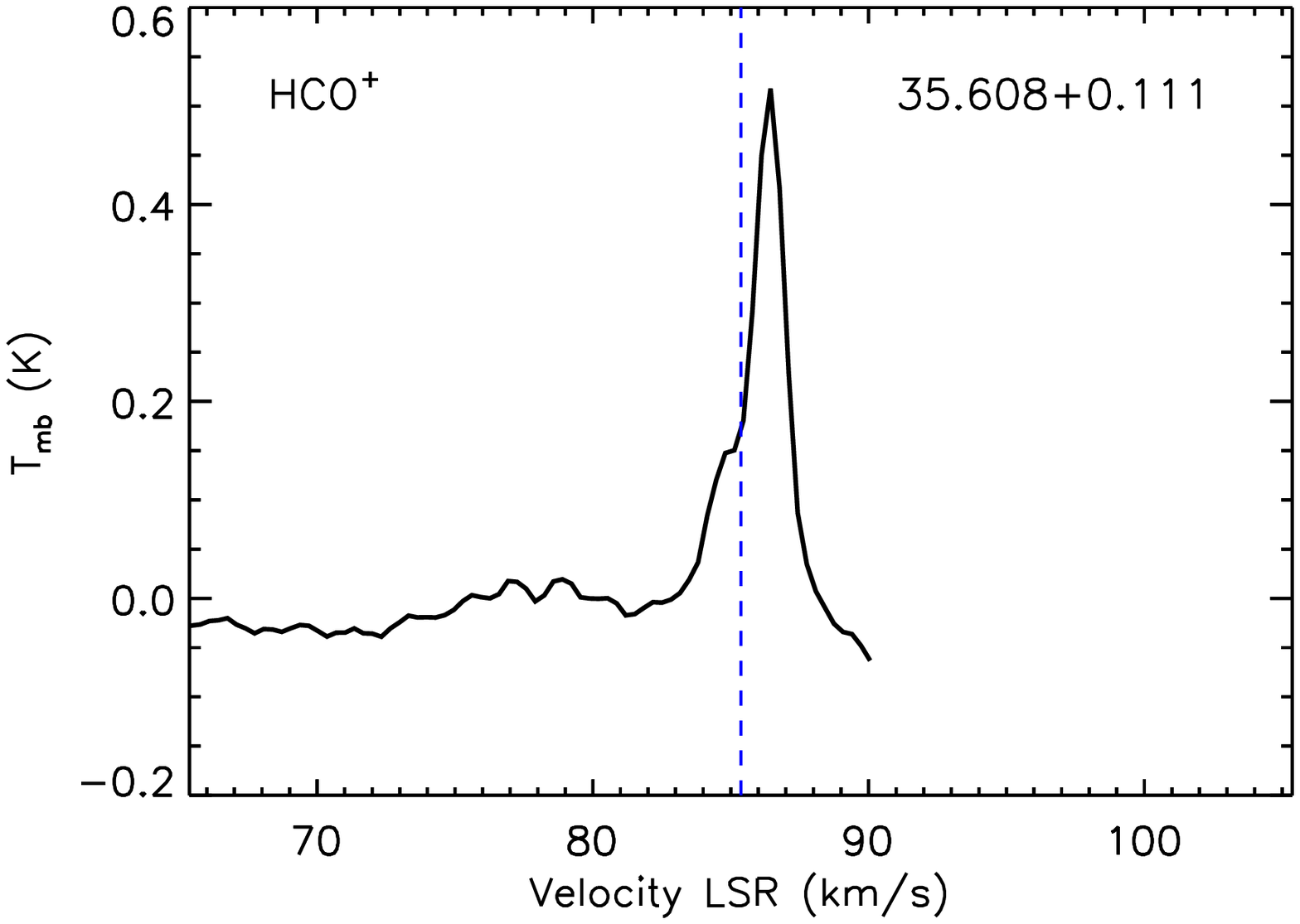}
\includegraphics[width=4.2cm]{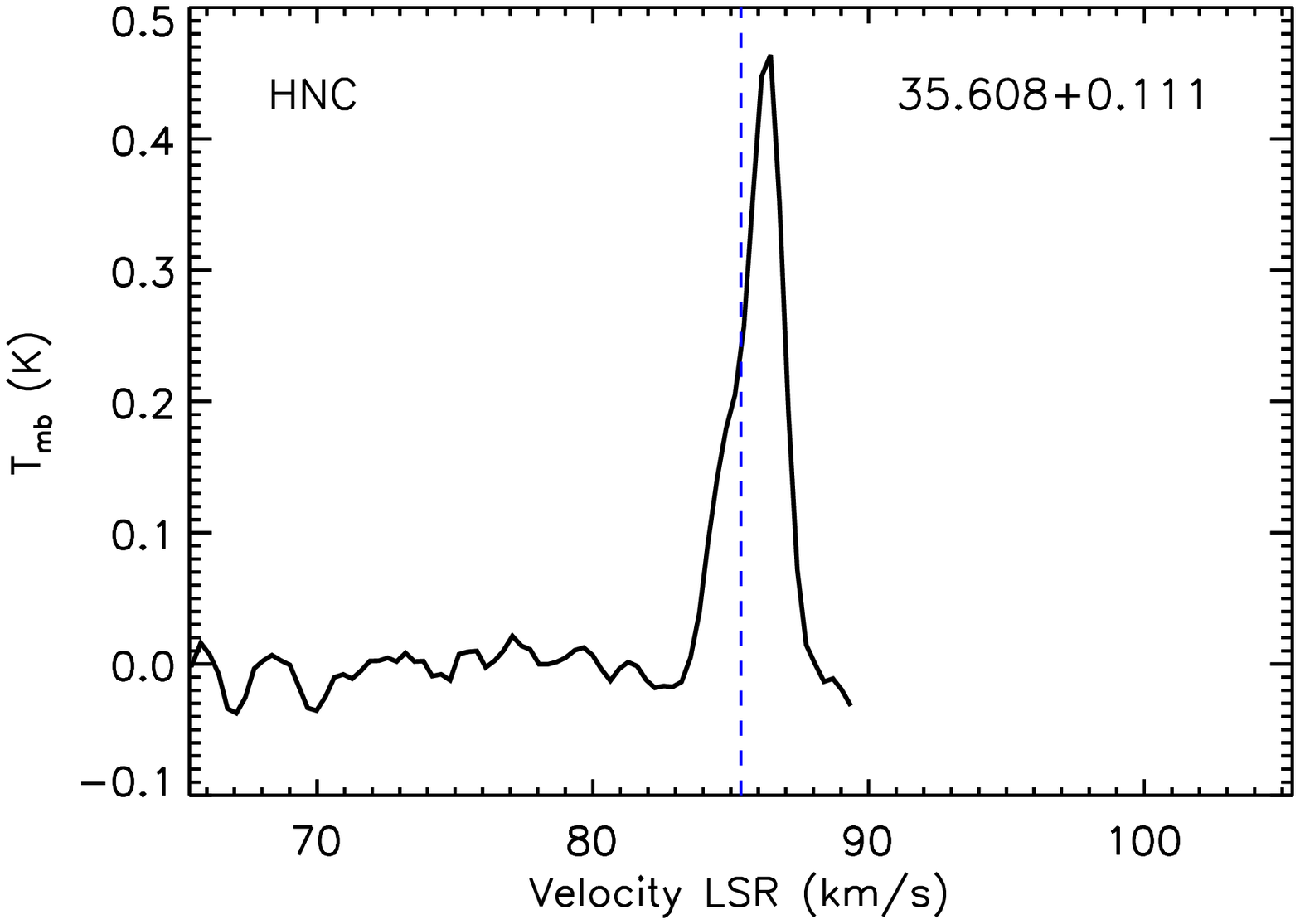}
\includegraphics[width=4.2cm]{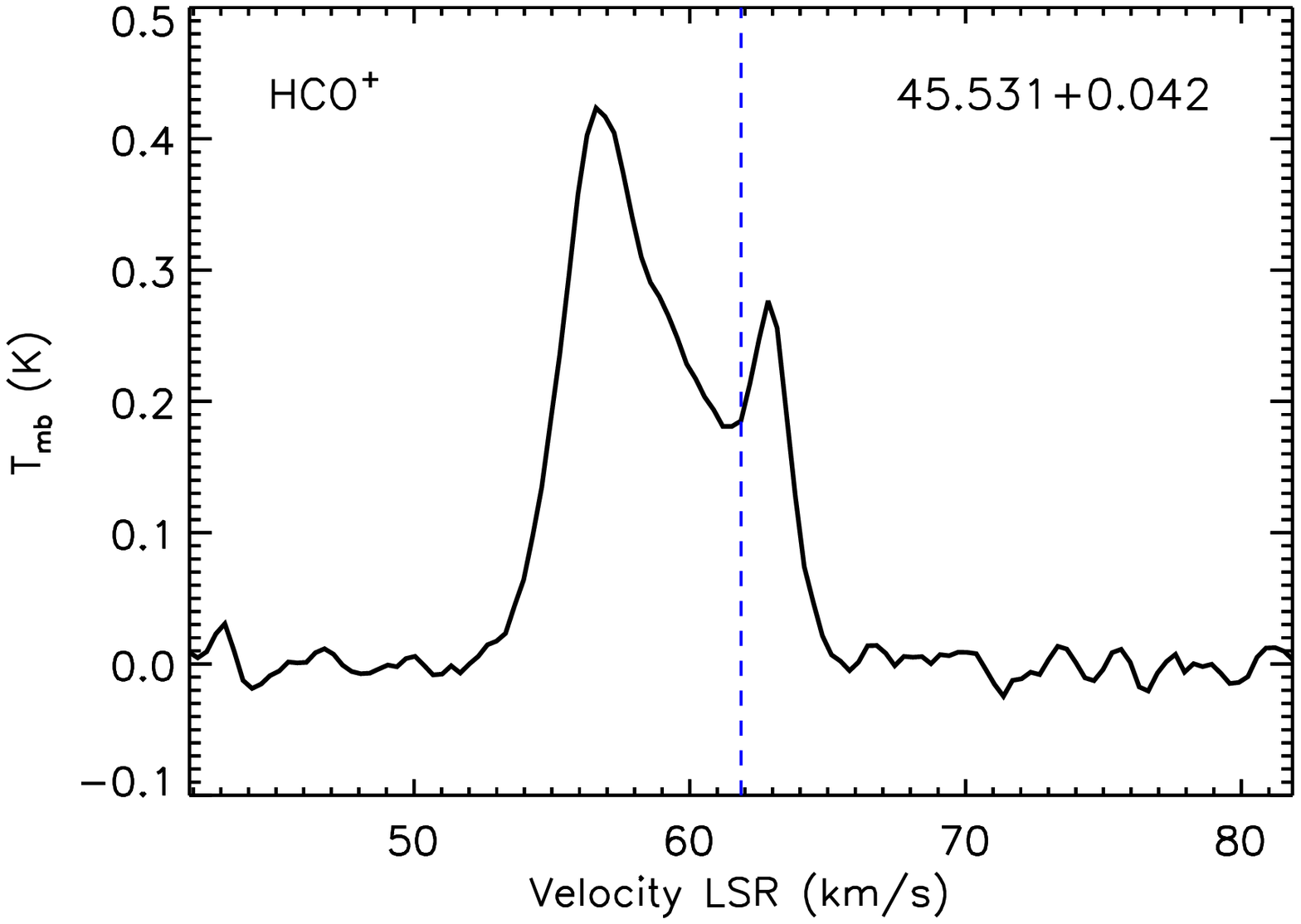}
\includegraphics[width=4.2cm]{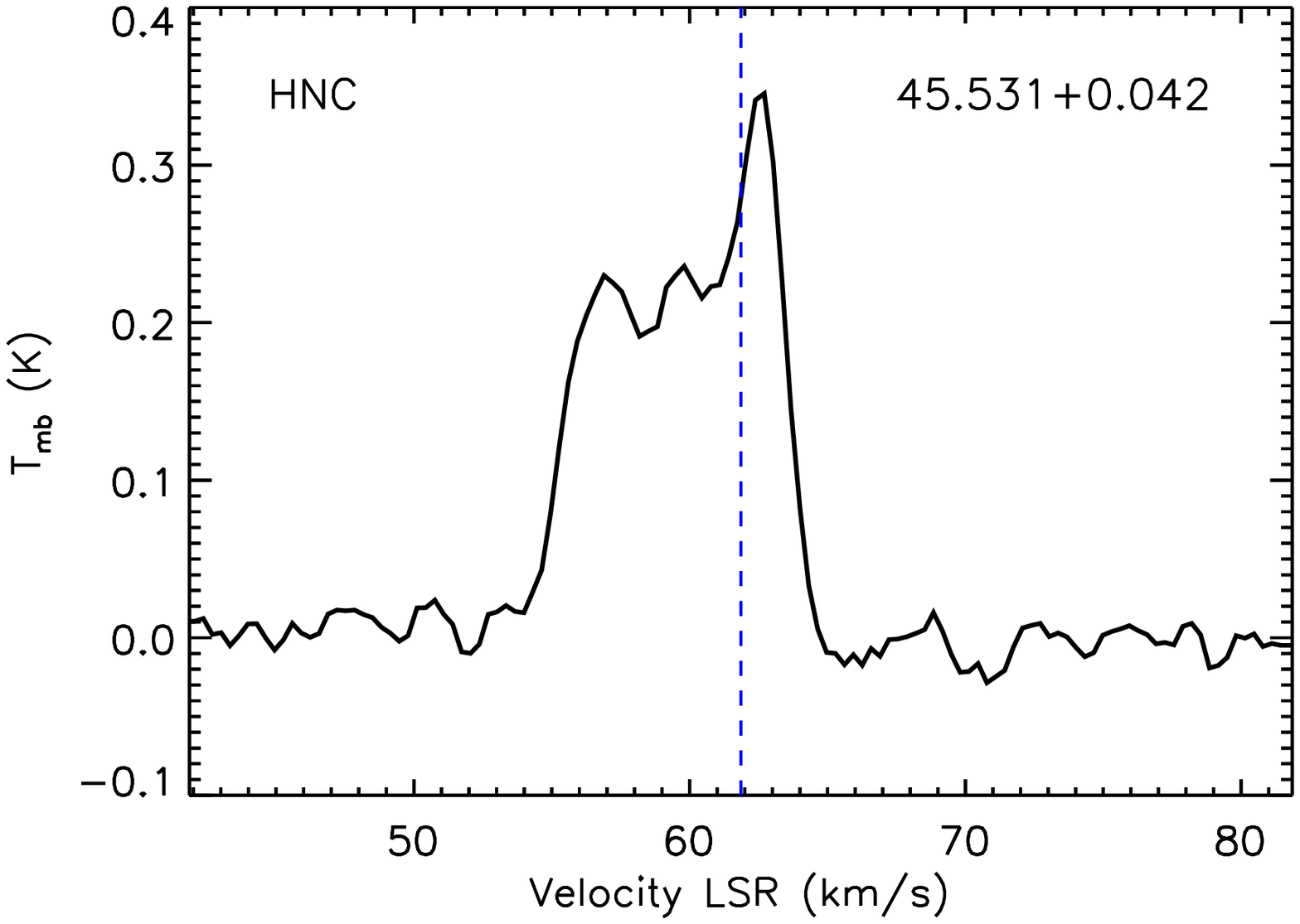}
\caption{\hco\ $(1-0)$ and HNC $(1-0)$ spectra averaged across the clump region for six representative clumps. Sources 30.131-0.644 and 49.732-0.471 have spectra symmetric around the central velocity of the clump defined by the \n2h\ $(1-0)$ emission (the blue-vertical line). Sources 24.552+0.096, 26.432-0.662, 35.608+0.111 and 45.531+0.042 have the \hco, the HNC or both asymmetric spectra with respect to the central velocity of the clump.}    
\label{fig:hco+_hnc_spectra}
\end{figure*}

We looked at this dynamical activity at the various surface density regimes searching for asymmetries in the \hco\ $(1-0)$ and HNC $(1-0)$ \citep{Kirk13} spectra in all our sources. The majority of the spectra in the $\Sigma_{low}$ and $\Sigma_{int}$ clumps are symmetric (two examples are given in the top row of Figure \ref{fig:hco+_hnc_spectra}), with the exception of 4 clumps which show hints of asymmetries in only the \hco\ $(1-0)$ spectrum (26.432-0.662) or in both \hco\ $(1-0)$ and HNC $(1-0)$ lines (24.552+0.096, 35.608+0.111 and 45.531+0.042). These spectra are shown in Figure \ref{fig:hco+_hnc_spectra}. The asymmetries are evident once we overplot the central velocity of the \n2h\ $(1-0)$ line that we have assumed as the optically thin line and used to estimate the central velocity of each region. The asymmetries in the clumps 35.608+0.111 and 26.432-0.662 can be due to the dynamical activity of the regions, while clumps 24.552+0.096 and 45.531+0.042 have more complex spectra and the interpretation is less obvious. 

The previous results suggest that, as moving towards denser and denser regions and approaching the critical surface density value of $\Sigma\simeq0.1$ g cm$^{-2}$, the contribution of the increasing gravitational potential starts to dominate over the turbulence in driving the observed dynamics at the clump scales.

\subsection{Filament analysis}\label{sec:results_clouds}
In this Section we investigate the properties of the natal filaments and their dynamics, separating them in the same three groups identified by the properties of their embedded 70\mum\ quiet clumps.

The filament masks derived from the Hi-GAL column density maps \citep{Schisano19} delimit the \co13\ $(1-0)$ regions that we extract to estimate the properties of each object. For each filament we take the length, mass and linear mass from the Hi-GAL filaments catalogue. We obtain the length $L$ of each filament from the angular length determined in the \citet{Schisano19} catalogue, and assuming the same distance of the embedded clump derived in \citet{Traficante15a}. In this way we also determined the length of two filaments in the Hi-GAL catalogue with no determined distance, 49.398-0.576 and 49.732+0.471. We have also derived the filament mass from the \co13\ GRS data, together with the central velocity, equivalent radius, and velocity dispersion parameters. The clump 45.531+0.042 partly overlaps with two separated filaments in the \citet{Schisano19} catalogue with the CO peak emission at very similar velocities. We have estimated the central velocity and the velocity dispersion of the cloud as the average values of the two filaments combined, but it was not possible to uniquely define the length and the linear mass of the combined cloud. For this object we did not consider these parameters in the analysis.

In order to evaluate the filament properties from the \co13\ datacubes, we need first to determine the central velocity $v_{LSR, fil}$ and the velocity range $v_{fil}$ that defines each filament. We initially fix the central velocity of each filament equal to the $v_{LSR}$ of the clump determined from the \n2h\ $(1-0)$ fit, $v_{LSR, cl}$, and we fix $v_{fil}$ equal to 10 km s$^{-1}$, defined as a $\pm5$ km s$^{-1}$ spectral window around $v_{LSR, cl}$. This range is enough to include the emission of each filament (the velocity gradient we measure is few km s$^{-1}$) and to exclude the contribution from other clouds that may be present along the line of sight.

\begin{figure*}
\centering
\includegraphics[width=10cm]{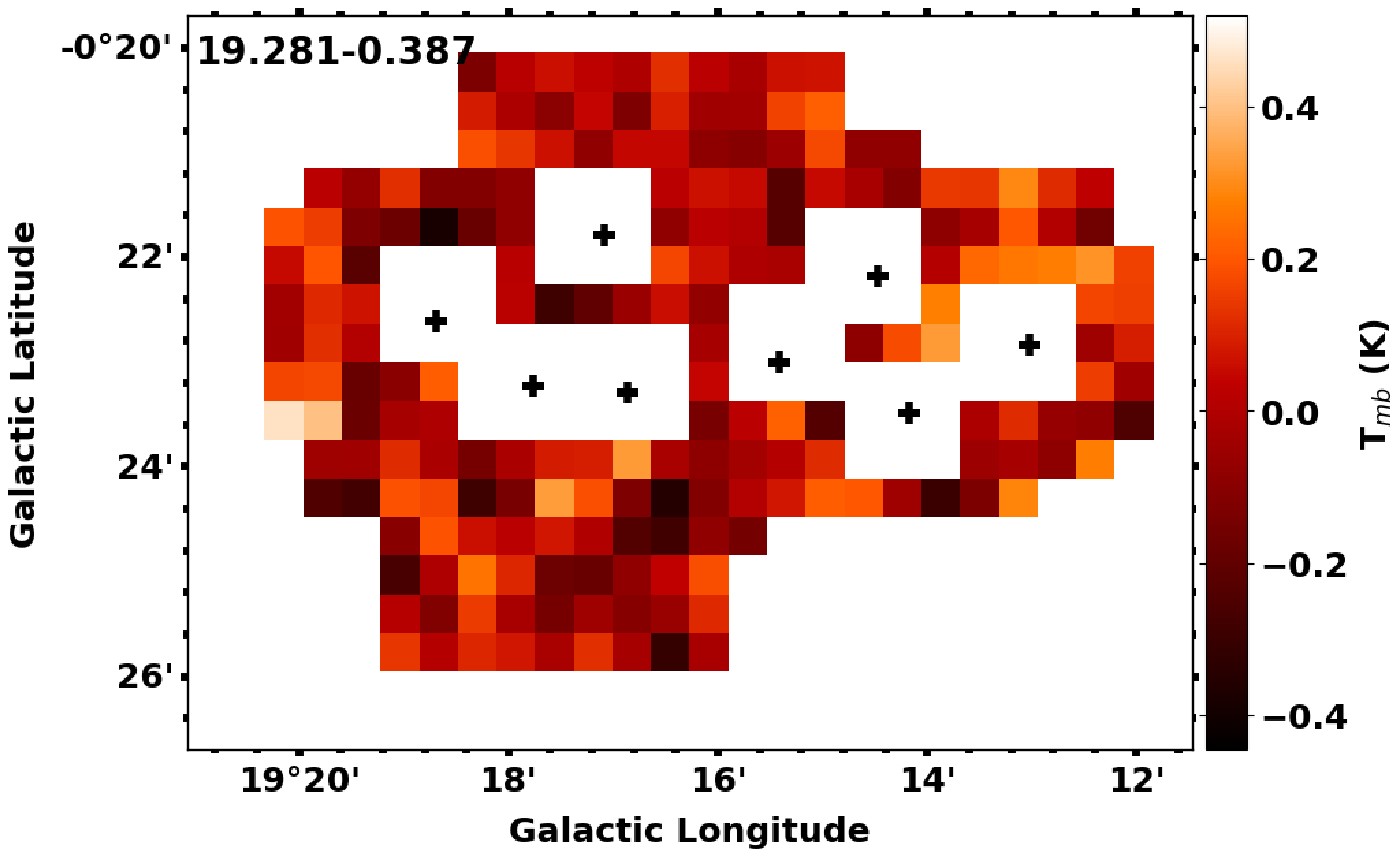} \qquad 
\includegraphics[width=6.8cm]{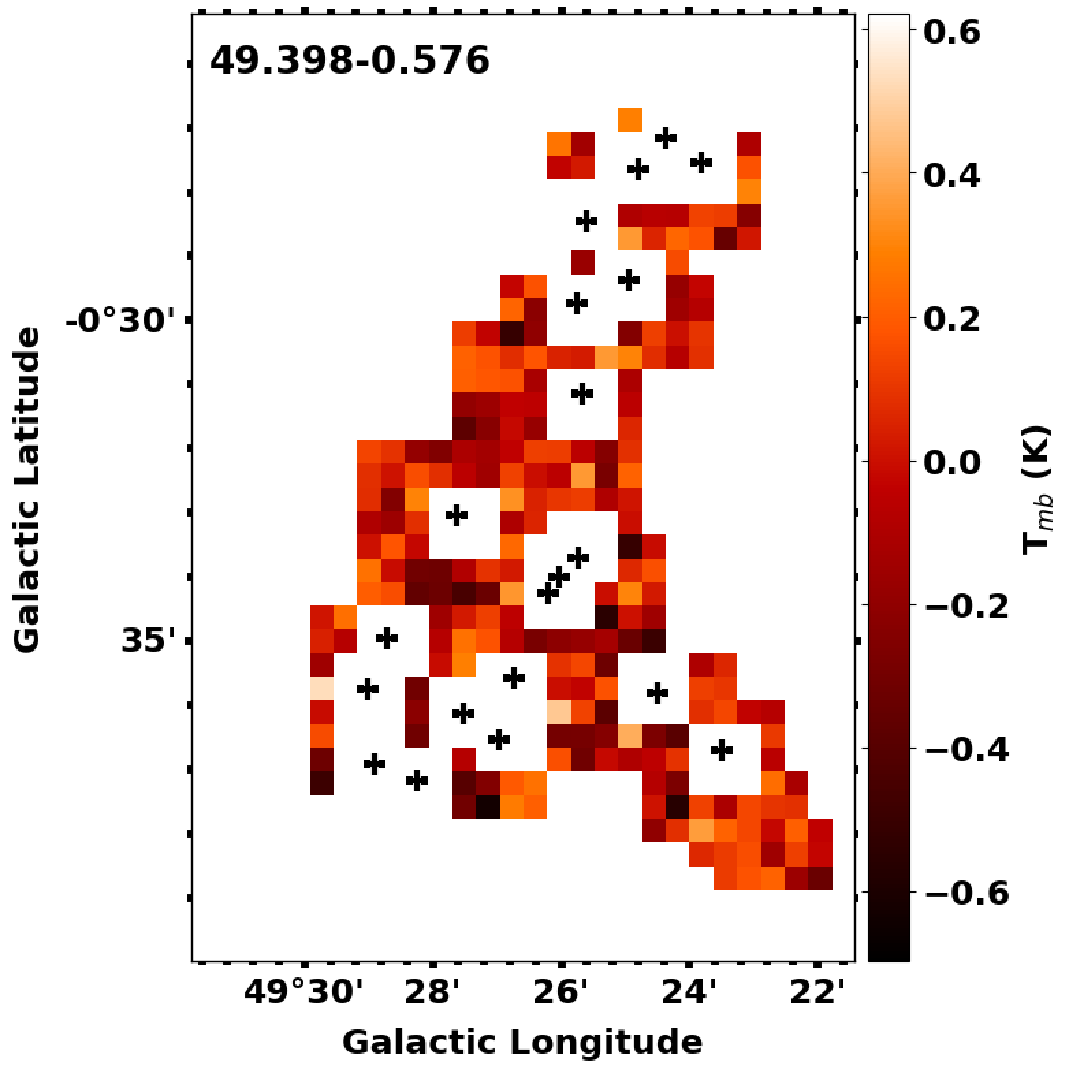}
\caption{A spectral channel of the \co13\ GRS data cube for the filaments 19.281-0.387 and 49.398-0.576 (the same of Figure \ref{fig:fil_filament_clump}). The black crosses are the centroids of the sources extracted from the \citet{Elia17} catalogue. A box of 3x3 pixels centered on each embedded Hi-GAL source has been masked to minimize the uncertainties of the filament kinematics across line of sights where the CO emission can be optically thick. The CO emission across the un-masked pixels is assumed to be optically thin and is used to estimate the kinematics of each filament.}    
\label{fig:fil_higal_sources_masked}
\end{figure*}

Before refining the calculation of $v_{LSR, fil}$ (and redetermining the extremes of $v_{fil}$), we need to take into account that the \co13\ $(1-0)$ emission can easily become optically thick along the line of sight of the dense clumps. To minimize the uncertainties derived from the optical depth effects, we masked in the GRS cubes a box of 3x3 pixels (to consider a GRS beam) centered on each clump identified in the \citet{Elia17} catalogue. The \co13\ emission in the remaining pixels is assumed to be optically thin and these pixels are used to estimate the central velocity and $v_{fil}$ of each region. The differences in the average values with and without including these lines of sight (LOS) are however relatively small, few percents on average. Two examples of the masked maps for the filaments 25.609+0.228 and 49.398-0.576 are in Figure \ref{fig:fil_higal_sources_masked}.

Starting from the first estimate of $v_{fil}$ and the masked cubes for each filament we have estimated $v_{LSR, fil}$ with an iterative approach, starting from the definition of $v_{LSR, fil}$ as the first moment of the \co13\ $(1-0)$ cubes \citep[e.g.][]{Roman-Duval10}:

\begin{equation}\label{eq:average_velocity_co}
v_{LSR, fil}=\frac{\sum_{l,b,v_{fil}}T_{mb}(l,b,v)v}{\sum_{l,b,v_{fil}}T_{mb}(l,b,v)}
\end{equation}

where the sum is done across all $(l,b)$ pixels contained within the filament region in the masked cubes and in a velocity range $v_{fil}$. We further consider only the pixels with $S/N>4$ at the peak position, where the noise level of the map is evaluated separately for each filament in a 30 km s$^{-1}$ spectral window along a line of sight with no \co13\ emission. 

The value of $v_{LSR, fil}$ we obtain from Equation \ref{eq:average_velocity_co} is then used to determine the new $\pm5$ km s$^{-1}$ $v_{fil}$ window across which re-evaluate $v_{LSR, fil}$. Usually, after the second iteration the value of $v_{LSR, fil}$ reaches convergence, and we stop the run and fix the value, otherwise we perform a third iteration and fix the value of $v_{LSR, fil}$ (and $v_{fil}$) at the end of this run. In this way we can also estimate the velocity shift between the central velocity of the clumps and the average central velocity of the natal filaments. This shift is on average of $\simeq1.2$ km s$^{-1}$, and for only 3 objects the difference is $> 2$ km s$^{-1}$, as shown in Table \ref{tab:filament_params}. 

Once we have defined $v_{LSR, fil}$ (and consequently the range $v_{fil}$), we can derive all the other parameters of the filaments from the CO datacubes: CO mass $M_{CO}$, linear mass $M_{line, CO}$, equivalent radius \req, average surface density and velocity dispersion. Only the latter is evaluated in the masked maps, since it may be affected by the optical depth effects. The other parameters are evaluated along all the good pixels, including the LOS of the Hi-GAL sources, because the mass estimation includes the effects of the optical depth, and the regions where the clumps are embedded contribute strongly to the physical parameters (mass, equivalent radius and surface density) of the filaments. The parameters derived for each filament are listed in Table \ref{tab:filament_params}.

The \co13\ mass $M_{CO}$ is derived following \citet{Roman-Duval10}:

\begin{equation}
M_{CO}=0.27\frac{\mathrm{d^{2}}}{\mathrm{kpc^{2}}}\int_{l,b,v_{fil}}\frac{T_{ex}(l,b,v)\tau_{13}(l,b,v)}{1-e^{\frac{-5.3}{T_{ex}(l,b,v)}}}\frac{dv}{\mathrm{km\ s^{-1}}}\frac{dl}{'}\frac{db}{'}
\end{equation}

where $\tau_{13}$ is the opacity and $T_{ex}$ is the excitation temperature evaluated for each voxel $(l,b,v)$ within the 10 km s$^{-1}$ interval defined by $v_{fil}$. The integral is in this case performed along all the good pixels, including the LOS of the Hi-GAL sources. The opacity in each voxel is estimated as

\begin{equation}
\tau_{13}(l,b,v)=-\mathrm{ln}\ \bigg(1-\frac{0.189\ T_{13}(l,b,v)}{(e^{\frac{5.3}{T_{ex}(l,b,v)}}-1)^{-1}-0.16}\bigg)
\end{equation}

\noindent where $T_{13}(l,b,v)$ is the \co13\ $(1-0)$ brightness temperature of each voxel. 

The \co13\ $(1-0)$ excitation temperature at each voxel is determined from the excitation temperature of the \12co\ $(1-0)$ line, under the assumption that these two temperatures are equal along our filaments. Assuming also that the \12co\ line is optically thick along the filaments, the excitation temperature at each voxel $T_{ex}(l,b,v)$ can be derived from the \12co\ brightness temperature as

\begin{equation}
T_{ex}(l,b,v)=5.53\frac{1}{\mathrm{ln}\ \big(1+\frac{5.53}{T_{12}(l,b,v)+0.837}\big)}
\end{equation}

The \12co\ emission along our filaments is taken from the FUGIN datacubes \citep{Umemoto17}. We first create a $T_{ex}$ cube from the \12co\ data at the FUGIN spatial resolution (20\arcsec), then we smoothed these maps to the GRS resolution (55\arcsec). 

\begin{figure}
\centering
\includegraphics[width=8.8cm]{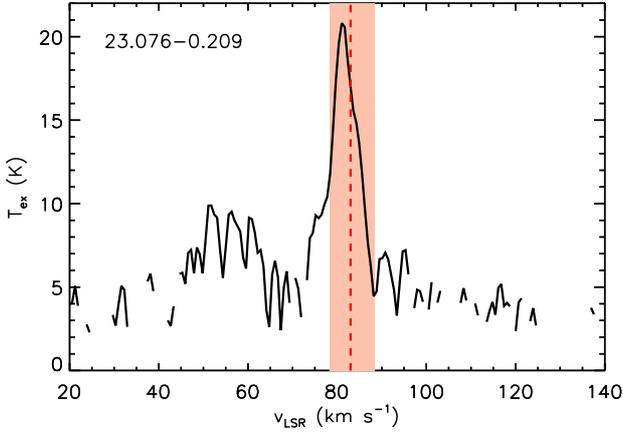}
\caption{Excitation temperature \tex\ along a pixel of the filament 23.076-0.209, estimated from the FUGIN data cubes and smoothed at the GRS resolution. The missing points are in correspondence of low $S/N$ pixels in the \12co\ FUGIN data cubes. The red-dotted vertical line is the central velocity of the filament $v_{LSR, fil}$ and the red-shaded area is the corresponding $v_{fil}$ estimated as discussed in the main text.}    
\label{fig:Tex_filament}
\end{figure}

The velocity resolution of the FUGIN cubes (0.65 km s$^{-1}$) is $\simeq3$ times lower than the GRS one ($\simeq$0.2 km s$^{-1}$). We derive the voxel cubes of $T_{ex}$ assuming at each GRS velocity channel the $T_{ex}$ of the closest FUGIN velocity channel. In this way, the $T_{ex}$ estimated in each FUGIN channel is associated, on average, to three GRS consecutive velocity channels. An example for a pixel of the source 23.076-0.209 is in Figure \ref{fig:Tex_filament}. Since the FUGIN survey does not cover the clouds above longitude $l=50\deg$, for the cloud associated with the clump 53.361+0.042 we assumed a fixed excitation temperature of $T_{ex}=10$ K, in agreement with the average value of the rest of the sample ($T_{ex,mean}=8.7\pm2.5$ K).

\begin{figure}
\centering
\includegraphics[width=8.8cm]{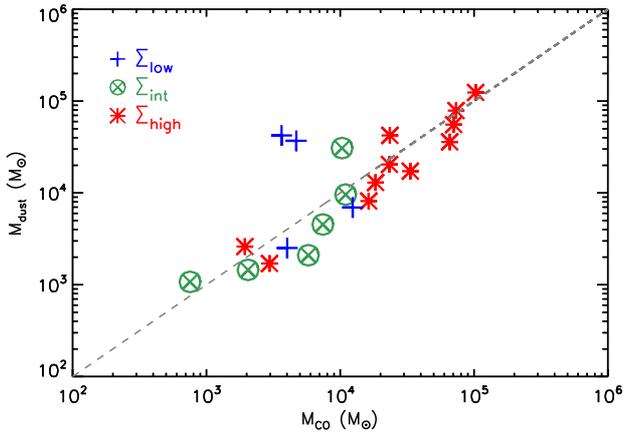}
\caption{Mass of the filaments derived from dust, $M_{dust}$, vs. mass derived from CO, $M_{CO}$. The black dashed line is the $y=x$ bisector. The values for the three groups discussed in Section \ref{sec:results} are represented with different symbols/colors. Objects associated with clumps in the $\Sigma_{low}$, $\Sigma_{int}$ and $\Sigma_{red}$ groups are represented as blue crosses, green circles with X and red asterisks respectively.}    
\label{fig:mass_fil_CO_mass_fil_dust}
\end{figure}

The mass derived from dust and from CO are compared in Figure \ref{fig:mass_fil_CO_mass_fil_dust}. There is a good correlation between the two (the black dashed line is the \textit{y=x} bisector). Two filaments, 28.19-0.192 and 36.608+0.111, have a significant higher dust mass compared with the CO mass, both belonging to the $\Sigma_{low}$ group. Both these filaments show multiple CO components along the line of sight, and the clump identified in the \n2h\ emission is not associated with the main CO peak. Therefore, the mass obtained from the integrated continuum infrared emission is likely overestimated for both filaments. Given the complexity of the CO spectrum along each line of sight, that often intercepts many clouds at likely different dust temperatures, in the following we will only consider the mass derived from the CO data.

The linear mass $M_{line, CO}$ is obtained as $M_{CO}/L$. For each filament we have also derived an equivalent radius \req, defined as \req=$\sqrt{A/\pi}$, with $A$ equal to the area of the \co13\ pixels with well-defined emission (including pixels along the line of sights of the Hi-GAL clumps). From the $M_{CO}$ and \req\ we determined the average surface density of each filament $\Sigma_{fil}=M_{CO}/(\pi R_{eq}^{2})$.

To study the filament kinematics we have derived the velocity dispersion from the CO data cubes in two different ways. First, we evaluated the second moment of the \co13\ $(1-0)$ maps, i.e. the intensity weighted velocity dispersion \citep{Roman-Duval10} in the $v_{fil}$ velocity range defined as:

\begin{equation}
\sigma_{gl}^{2}=\frac{\sum_{l,b,v_{fil}}T_{mb}(l,b,v)(v-v_{LSR, fil})^{2}}{\sum_{l,b,v_{fil}}T_{mb}(l,b,v)}
\end{equation}

This is the average velocity dispersion of the whole filament, including the contribution of the large-scale velocity gradients that broaden the width of the average spectrum and have several different origins: they can originate from the dynamics of the gas, like e.g. large-scale infall motions \citep{Kirk13} or gravitational collapse along the filament \citep{Peretto14}. They may be due to large-scale motions not related with the internal dynamics of the gas, such as rotation \citep{Kirk13} or compression from an external source \citep{Williams18}, and, for the most elongated objects, they may also originate from Galactic scale motions such as Galactic rotation, shear or compression \citep[e.g.][]{Duarte-Cabral16}. Alternatively, they may have a non-physical origin and be the resultant from the blending of the emission from unresolved sub-structures that may broaden the observed width \citep{Beuther13} and in some cases mimic a large-scale gradient \citep{Henshaw14}.

In order to consider the motions in the filaments without these large-scale gradients, we also derived the velocity dispersion of each filament without including their contribution. To do this, we have first derived the map of the second moment per pixel, defined as:

\begin{equation}\label{eq:velocity_dispersion_local_map}
\sigma_{loc}^{2}(l,b)=\frac{\sum_{v_{fil,loc}}T_{mb}(l,b,v)(v(l,b)-\langle v_{loc}(l,b)\rangle)^{2}}{\sum_{v_{fil,loc}}T_{mb}(l,b,v)}
\end{equation}

\noindent where $\langle v_{loc}(l,b)\rangle$ is equivalent to the central velocity $v_{LSR, fil}$ defined in Equation \ref{eq:average_velocity_co} but now evaluated separately for each pixel. Since $v_{fil,loc}(l,b)$ is estimated separately for each pixel, effectively treating all of them as independent pixels, we can get an estimate of the average non-thermal motions per filament without large-scales contribution, $\langle\sigma_{loc}\rangle$, by evaluating the mean value of $\sigma_{loc}(l,b)$. 

\begin{figure}
\centering
\includegraphics[width=8.8cm]{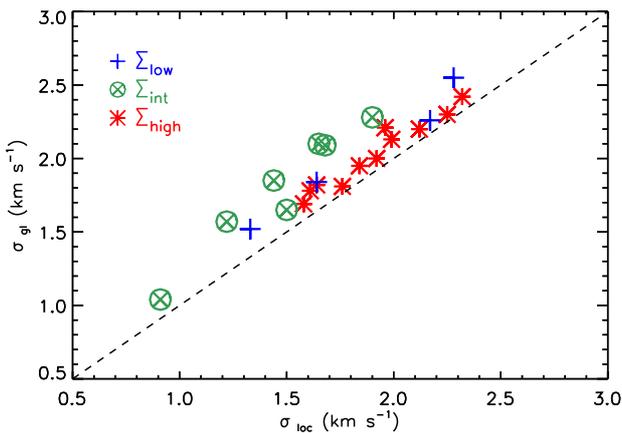}
\caption{Velocity dispersion of the filaments derived including the large scale gradients, $\sigma_{gl}$, vs. velocity dispersion derived after the removal of these gradients, $\sigma_{loc}$. The results for the clump 28.19-0.192 are not included since the $\sigma_{gl}$ value is significantly larger than $\sigma_{loc}$ because of a second CO component along the line of sight, as explained in Section \ref{sec:results_clouds}, and it goes outside the plotted ranges.}
\label{fig:velo_fil_gl_velo_fil_loc}
\end{figure}

The values of $\sigma_{gl}$ and $\sigma_{loc}$ and their ratio are in Table \ref{tab:filament_params}, and in Figure \ref{fig:velo_fil_gl_velo_fil_loc} we show $\sigma_{loc}$ against $\sigma_{gl}$ for each filament. The values of $\sigma_{loc}$ are systematically lower than $\sigma_{gl}$, and the difference between them provides also an indication of the contribution of the large-scale kinematics to the measured non-thermal motions in each cloud. These differences are more pronounced in the $\Sigma_{low}$ and $\Sigma_{int}$ clumps rather than in $\Sigma_{high}$ clumps, and they are of the order of $simeq30\%$ at most. The only exception if for the filament 28.19-0.192, where the difference is larger than 100\%. This is due to the contribution of a second CO component along the line of sight that contributes to the estimate of $\sigma_{gl}$, and it is likely the same component that also contributed to the over-estimateion of the mass of this filament from the dust emission maps showed in Figure \ref{fig:mass_fil_CO_mass_fil_dust}.

These relatively small differences between $\sigma_{loc}$ and $\sigma_{gl}$ imply that the observed motions are dominated by the internal motions of the cloud, and not by the mechanism(s) responsible for the observed large-scale gradients.

\begin{center}
\begin{table*}
\centering
\begin{tabular}{c|c|c|c|c|c|c|c|c|c|c|c|c|c}
\hline
\hline
 Clump & $M_{dust}$ & $M_{CO}$ & $L$ & $M/L$ & $M_{CO}/L$ & $R_{eq}$ & $\Sigma_{fil}$ & v$_{LSR, cl}$ & v$_{LSR, fil}$ & $\sigma_{loc}$ & $\sigma_{gl}$ & $\sigma_{ratio}$ & Group \\
  & (M\sun) & (M\sun) & (pc) & (M\sun\ pc$^{-1}$) & (M\sun\ pc$^{-1}$) & (pc) & (g cm$^{-2}$) & (km s$^{-1}$) & (km s$^{-1}$) & (km s$^{-1}$) &   (km s$^{-1}$) & & \\
\hline

       49.732-0.471  &$--$  &       12837  &    17.4  &        $--$  &       737.2  &     5.7  &   0.027  &   68.51  &   67.62  &    1.33  &    1.52  &    1.14  &  $\Sigma_{low}$ \\
        28.19-0.192  &       42320  &        3648  &    25.9  &      1632.3  &       140.7  &     4.7  &   0.011  &  100.70  &   99.44  &    1.54  &    3.24  &    2.10  &  $\Sigma_{low}$ \\
       45.531+0.042  &        $--$  &       $--$  &     $--$  &       $--$  &      $--$  &     $--$  &  $--$  & 61.86  &   58.54  &    2.02  &    2.15  &    1.06  &  $\Sigma_{low}$ \\
       23.076-0.209  &        2516  &        4020  &     5.7  &       440.8  &       704.2  &     2.3  &   0.051  &   82.66  &   83.36  &    2.17  &    2.26  &    1.04  &  $\Sigma_{low}$ \\
       26.432-0.662  &        6946  &       12405  &    15.2  &       458.3  &       818.6  &     4.1  &   0.048  &   62.38  &   63.23  &    1.64  &    1.84  &    1.12  &  $\Sigma_{low}$ \\
       35.608+0.111  &       36978  &        4699  &    36.6  &      1009.1  &       128.2  &     4.4  &   0.016  &   85.37  &   81.58  &    2.28  &    2.55  &    1.12  &  $\Sigma_{low}$ \\
       
& & & & & & & & & & & & \\
       
       30.357-0.837  &        4532  &        7421  &    16.8  &       270.0  &       442.1  &     4.9  &   0.021  &   78.80  &   76.37  &    1.65  &    2.10  &    1.27  &  $\Sigma_{int}$ \\
       53.361+0.042  &       30788  &       10326  &    57.7  &       533.3  &       178.9  &     4.2  &   0.038  &   22.66  &   23.48  &    0.91  &    1.04  &    1.14  &  $\Sigma_{int}$ \\
       30.131-0.644  &        1453  &        2049  &     5.5  &       264.9  &       373.5  &     2.6  &   0.020  &   79.81  &   81.28  &    1.44  &    1.85  &    1.28  &  $\Sigma_{int}$ \\
       15.631-0.377  &        1075  &         755  &     2.8  &       390.3  &       273.9  &     2.0  &   0.013  &   40.00  &   39.34  &    1.22  &    1.57  &    1.29  &  $\Sigma_{int}$ \\
       49.398-0.576  &$--$  &       19968  &    24.1  &        $--$  &       827.2  &     6.5  &   0.032  &   60.79  &   58.29  &    1.68  &    2.09  &    1.24  &  $\Sigma_{int}$ \\
       28.792+0.141  &        9601  &       10988  &    17.5  &       549.0  &       628.3  &     4.8  &   0.031  &  107.20  &  107.96  &    1.90  &    2.28  &    1.20  &  $\Sigma_{int}$ \\
       25.982-0.056  &        2098  &        5780  &     5.5  &       378.2  &      1042.0  &     3.0  &   0.044  &   89.80  &   90.25  &    1.50  &    1.65  &    1.10  &  $\Sigma_{int}$ \\
       
& & & & & & & & & & & & \\
       
       19.281-0.387  &        2603  &        1940  &     9.0  &       289.6  &       215.8  &     3.7  &   0.009  &   53.50  &   52.28  &    1.58  &    1.69  &    1.07  & $\Sigma_{high}$ \\
       34.131+0.075  &        1704  &        2963  &     3.2  &       540.1  &       938.8  &     1.9  &   0.055  &   56.80  &   57.17  &    1.76  &    1.81  &    1.03  & $\Sigma_{high}$ \\
       23.271-0.263  &      124147  &      103054  &    48.0  &      2587.8  &      2148.1  &     9.3  &   0.079  &   82.30  &   79.41  &    2.32  &    2.42  &    1.04  & $\Sigma_{high}$ \\
       31.946+0.076  &       55491  &       70580  &    43.6  &      1272.6  &      1618.7  &     8.1  &   0.071  &   96.40  &   96.17  &    1.84  &    1.95  &    1.06  & $\Sigma_{high}$ \\
       22.756-0.284  &        8159  &       16318  &    18.8  &       434.6  &       869.2  &     4.4  &   0.057  &  105.00  &  106.18  &    1.96  &    2.21  &    1.13  & $\Sigma_{high}$ \\
        22.53-0.192  &       35907  &       65881  &    32.2  &      1114.3  &      2044.5  &     7.7  &   0.074  &   76.20  &   77.02  &    2.25  &    2.30  &    1.02  & $\Sigma_{high}$ \\
       28.537-0.277  &       42291  &       23505  &    13.1  &      3222.6  &      1791.1  &     4.4  &   0.081  &   88.30  &   86.95  &    1.61  &    1.78  &    1.11  & $\Sigma_{high}$ \\
       28.178-0.091  &       17266  &       33436  &    28.0  &       616.6  &      1194.2  &     5.8  &   0.067  &   98.20  &   97.69  &    1.99  &    2.13  &    1.07  & $\Sigma_{high}$ \\
       24.013+0.488  &       78849  &       73495  &    88.9  &       887.2  &       827.1  &    11.6  &   0.036  &   95.00  &   95.70  &    1.64  &    1.82  &    1.11  & $\Sigma_{high}$ \\
       25.609+0.228  &       12977  &       18370  &    20.8  &       623.1  &       882.1  &     5.6  &   0.039  &  113.60  &  113.10  &    2.12  &    2.20  &    1.04  & $\Sigma_{high}$ \\
       18.787-0.286  &       20426  &       23354  &    15.3  &      1336.2  &      1527.6  &     4.5  &   0.077  &   65.70  &   65.10  &    1.92  &    2.00  &    1.04  & $\Sigma_{high}$ \\

\hline
\end{tabular}
\caption{Physical and kinematic properties of the filaments associated with the Hi-GAL clumps. Col. 1: name of the embedded clump; Cols. 2-3: mass of the filament derived from the dust emission and from the CO, respectively; Col. 4: length of the filament obtained from the \citet{Schisano19} catalogue and re-scaled assuming the distance of the clump; Cols. 5-6: linear mass of the filament derived from the dust mass and the CO mass respectively; Col. 7: equivalent radius of the filament; Col. 8: Central velocity of the clump determined from the hyperfine fitting of the \n2h\ $(1-0)$ emission; Col. 9: Central velocity of the filament determined as discussed in Section \ref{sec:results_clouds}; Cols. 10-11: velocity dispersion of the filament obtained without and with the contribution of the large-scale gradients, as discussed in Section \ref{sec:results_clouds}; Col. 12: ratio of the velocity dispersion obtained with and without the contribution of the large-scale gradients, $\sigma_{ratio}=\sigma_{gl}/\sigma_{loc}$; Col. 13: reference group.}
\label{tab:filament_params}
\end{table*}
\end{center}

\subsubsection{The dynamics in filaments}
Similarly to what we have done for the clumps, we analyze here some relations useful to investigate the dynamics of the filaments.

\begin{figure}
\centering
\includegraphics[width=8.2cm]{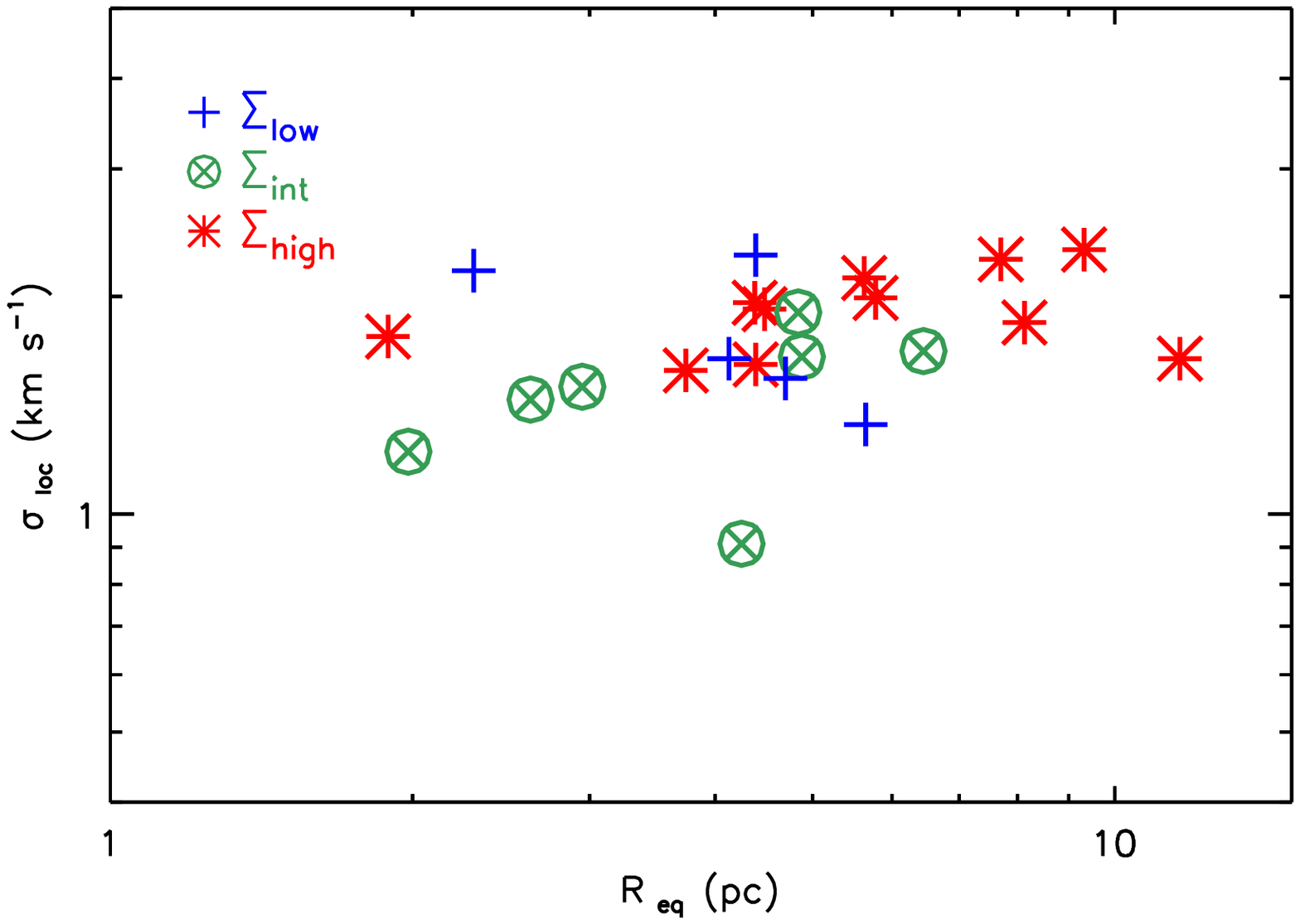}
\includegraphics[width=8.2cm]{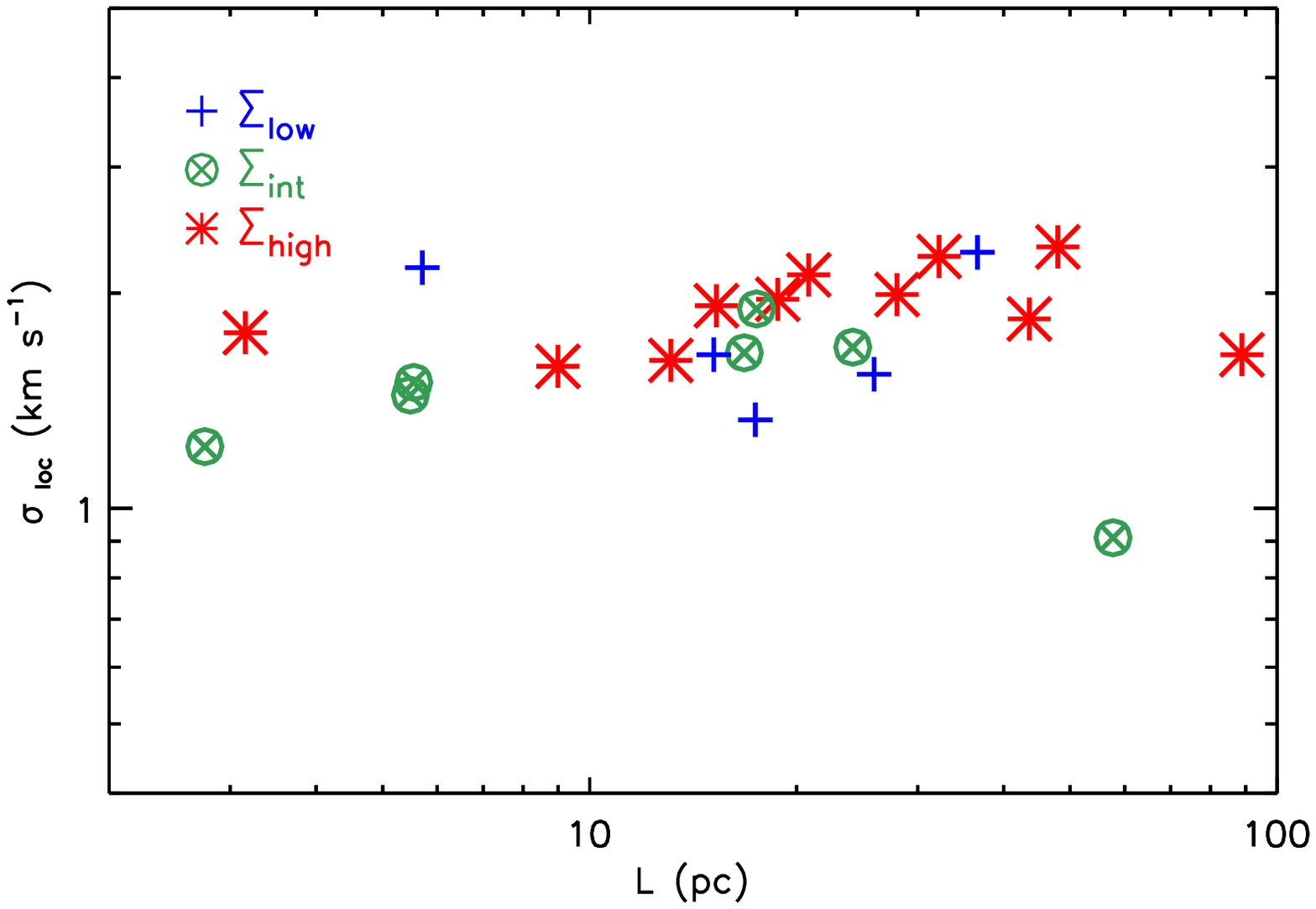}
\caption{\textit{Upper panel}: Larson relation evaluated for the filaments. The filaments size is considered as their equivalent radius $R_{eq}$. \textit{Lower panel}: same as before, but the size of the filament is equal to its length.}    
\label{fig:Larson_filaments}
\end{figure}

In order to evaluate the size-linewidth Larson's relation for our filaments, we have considered two possible definitions of the filament size. In Figure \ref{fig:Larson_filaments}, upper panel, we show the relation where the size of the filament is defined from its equivalent radius $R_{eq}$. However, these objects are particularly elongated, and an equivalent radius, although a useful parameter to define the size of a structure, may not be representative of the larger scales that may contribute to drive the non-thermal motions. The largest scale at which the kinetic energy can be injected in a filament is its whole length, $L$, and this is the other quantity that we have considered to derive the Larson relation (Figure \ref{fig:Larson_filaments}, lower panel). In both these diagrams there is no evident correlation between size and velocity dispersion, and this is true for all the filaments.

\begin{figure}
\centering
\includegraphics[width=8.2cm]{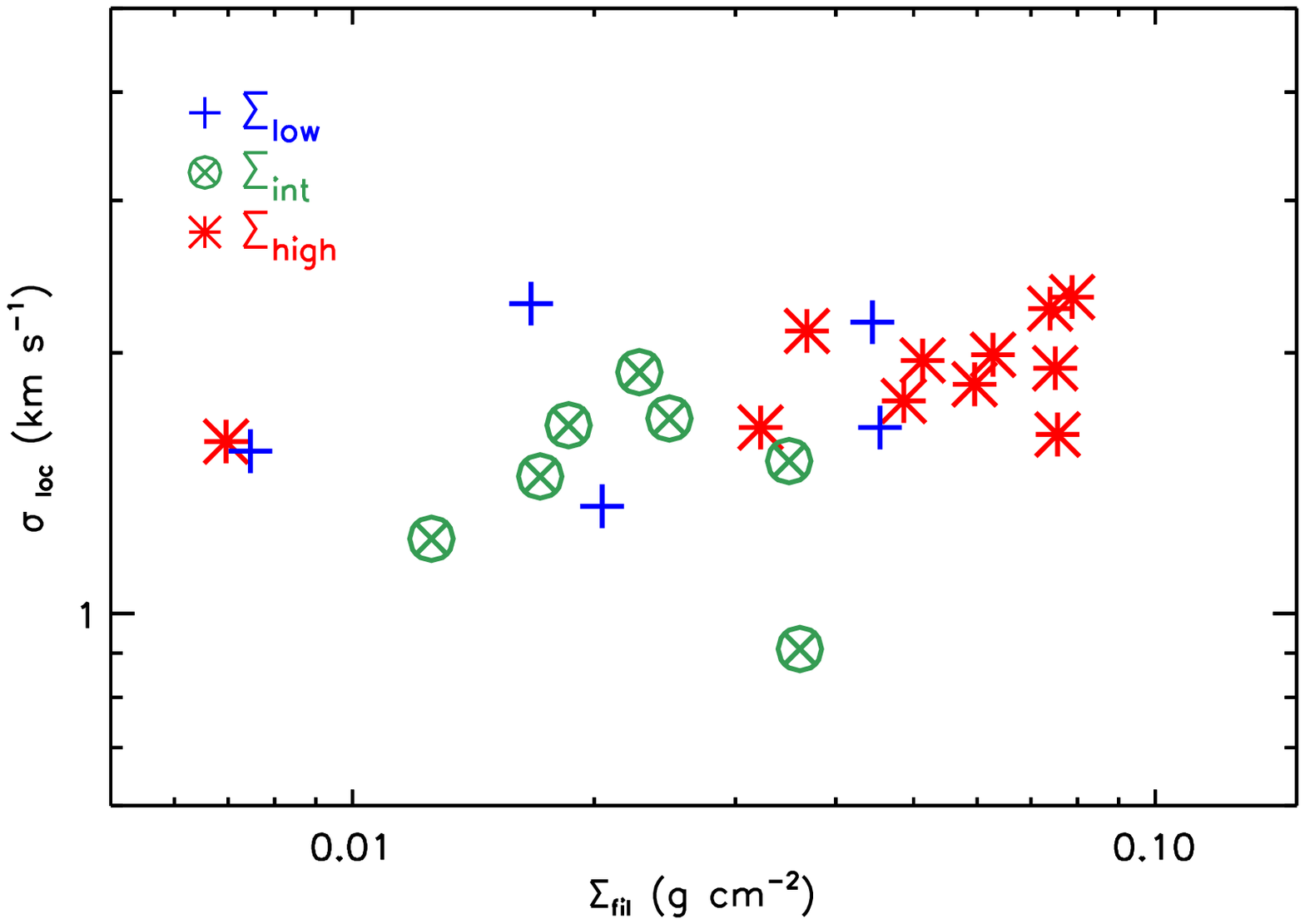}
\includegraphics[width=8.2cm]{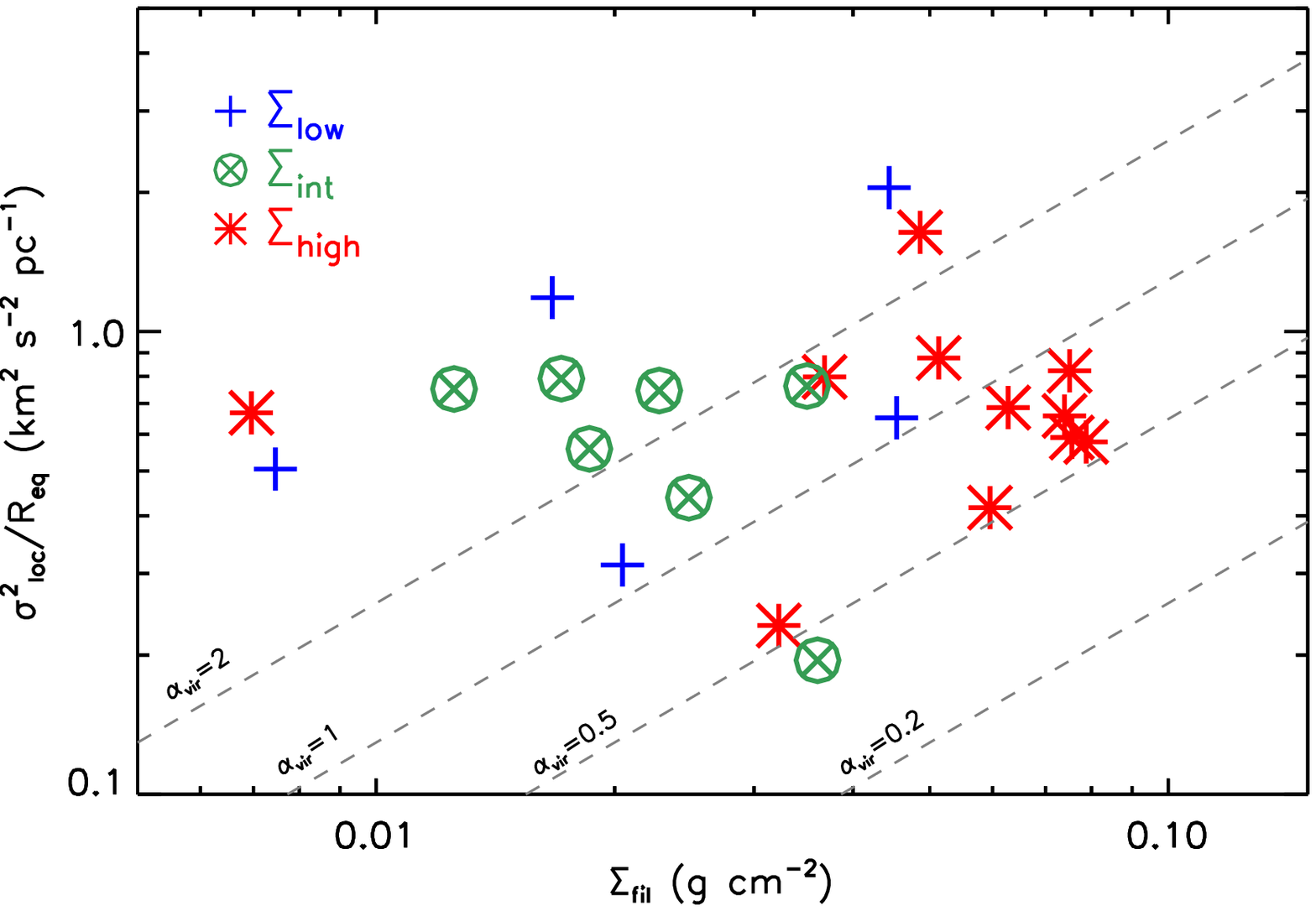}
\caption{\textit{Upper panel}: velocity dispersion of the filament (estimated after the removal of the large-gradient contribution) as function of the surface density of the filaments. \textit{Lower panel}: generalized Larson relation for the filaments. The grey-dashed lines are in correspondence of constant values of the virial parameter.}    
\label{fig:surf_dens_fil_velo_fil}
\end{figure}

In Figure \ref{fig:surf_dens_fil_velo_fil}, upper panel, we show the relation between the velocity dispersion $\sigma_{loc}$ and the surface density of the filaments. The correlation is very poor in all the groups, and this lack of correlation is also present in the generalized Larson relation, in the lower panel of Figure \ref{fig:surf_dens_fil_velo_fil}.

The lack of correlation in the previous diagrams could indicate that the average kinetic energy of each filament is determined by the energetics of the ambient environment, which can differ from region to region. Instead, in the next Section we will show how, at the filament-to-clump scales, the turbulent cascade is the most likely explanation for the observed non-thermal motions within each filament.

The filamentary structures themselves can accrete from the ambient gas and the inflow of new material contributes to generate a cascade of energy down to the star-forming sites that resembles the turbulent cascade, the so-called accretion-driven mechanism \citep{Klessen10}. In this case, filaments must evolve with time with the accretion of their mass per unit length $M_{line}$ and, correspondingly, their internal velocity dispersion. This scenario is observed in filaments detected in the \textit{Herschel} Gould Belt Survey \citep{Arzoumanian11}, with a power-law relation between $M_{line}$ and $\sigma$ of the form $M_{line}\propto\sigma^{0.31}$ \citep{Arzoumanian13}, although they are different objects compared with the ones analyzed in this work. \citet{Arzoumanian13} observed a collection of nearby, relatively low-mass filaments with sizes $\leq1$ pc, while the objects in the \citet{Schisano19} catalogue are distributed across the whole Galaxy and they can reach lengths up to tens of parsecs. 

\begin{figure}
\centering
\includegraphics[width=8.4cm]{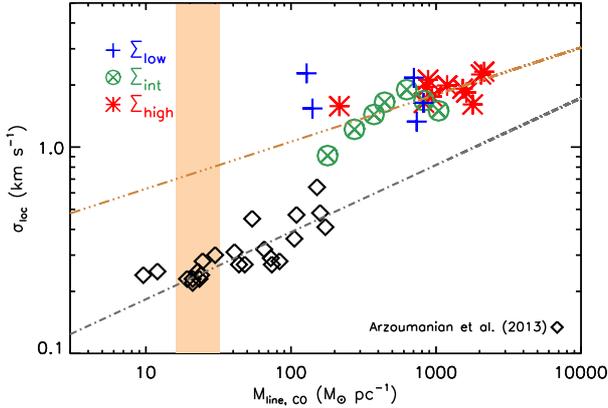}
\caption{Velocity dispersion of the filaments estimated without the contribution of the large-scales gradients as function of the linear mass derived from the CO data. The orange dash-dotted line is the fit to the filaments with clumps in the $\Sigma_{int}+\Sigma_{high}$ groups. The black open diamonds are the nearby filaments studied in \citet{Arzoumanian13}. The orange rectangle delimits the region above which the filaments are considered thermally super-critical, around the critical value of $M/L\simeq16$ M\sun/pc.}    
\label{fig:mass_length_sigma}
\end{figure}

To investigate if a similar mechanism can be responsible for the dynamics observed in our parsec-scales filaments, in Figure \ref{fig:mass_length_sigma} we plot the linear mass $M_{line, CO}$ as function of the velocity dispersion $\sigma_{loc}$ for our objects, superimposed with the results of nearby regions taken from \citet{Arzoumanian13}.

First, note that there is an offset between the velocity dispersion measured in nearby regions and in our filaments. This is likely to be attributed to the different gas tracers used to measure the motions. In the work of \citet{Arzoumanian13} the gas tracers are C$^{18}$O $(2-1)$ and \n2h\ $(1-0)$, both with higher critical density than the \co13\ $(1-0)$ used in this work and more suited to trace only the inner, denser regions of the filament.  

The correlation of the filaments in the $\Sigma_{high}$ group alone is mild (Pearson's coefficient is 0.56), and it increases if we combine the filaments in the $\Sigma_{int+high}$ groups ($\rho=0.71$). The fit to the data of these two groups together gives a relation $M_{line}\propto\sigma^{0.23}$, a shallower slope but still comparable with the results obtained in \citet{Arzoumanian13}. The few points of the regions belonging to the $\Sigma_{low}$ group are uncorrelated in this diagram. 

The same mechanism that drives the motion in nearby filaments could also drive the non-thermal motions observed at the scales of our filaments, but more data are needed to investigate this relation in a statistically significant way.


\section{The correlation between clumps and filaments}\label{sec:clumps_clouds_dynamics}
In this section we combine the results obtained separately for clumps and filaments to attempt a multi-scale analysis of our data.

\begin{figure}
\centering
\includegraphics[width=8.4cm]{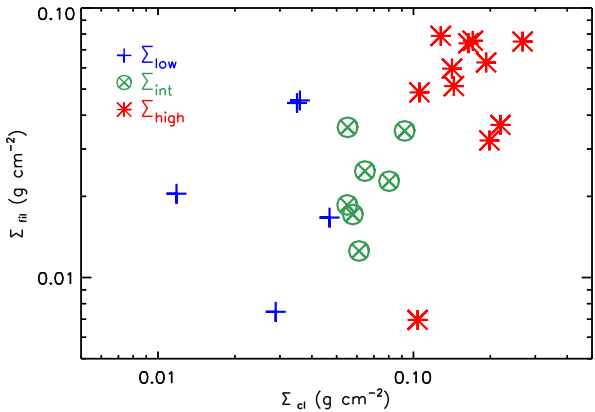}
\caption{Average surface density of the filament $\Sigma_{fil}$ versus the surface density of each clump $\Sigma_{cl}$ for the three groups as discussed in Section \ref{sec:results}.}    
\label{fig:surf_dens_fil_surf_dens_clump}
\end{figure}

The relation between the surface density of the clumps $\Sigma_{cl}$, and the average surface density of the parent filament, $\Sigma_{fil}$, is shown in Figure \ref{fig:surf_dens_fil_surf_dens_clump}. On average, the densest clumps in our sample are embedded in the filaments with the highest surface density (with the evident exception of 19.281-0.387 where the massive clump is embedded in a particularly low density filament). The large scatter observed for the values of the low density clumps is likely to be due to an observational bias. To show that, we compared the properties of our clumps with the properties of the other clumps embedded in each filament, derived from the \citet{Elia17} Hi-GAL clumps catalogue (the parameters of each clump have been re-scaled consistently at the distance of each filament). All our filaments embed more than one clump, the large majority of them being low and intermediate density clumps. The bias arises because 5/7 and 6/11 of the clumps identified in the $\Sigma_{int}$ and $\Sigma_{high}$ groups are the most massive ones embedded in their corresponding filaments, and they are the main responsible for the observed correlations. In contrast, all the filaments belonging to the $\Sigma_{low}$ group embed clumps more massive than the 70\mum\ quiet one analyzed here. We will come back to this point later in this section. 

This result is a further indication that the environment plays a significant role in the formation of massive stars: densest and more massive clumps seems to form preferentially along massive, dense filaments.

Next, we investigated the correlation between the kinematics of the clumps and the parent filaments. This can be done without any bias since we are tracing the kinematics of the clumps and filaments separately with different gas tracers. The \co13\ $(1-0)$ line, with a critical density of $\simeq3\times10^{3}$ cm$^{-3}$, is a good tracer of the relatively low column density material that delimits the filament, as opposed to the \n2h $(1-0)$ that, with a critical density of $\simeq6\times10^{4}$ cm$^{-3}$ \cite[and an effective critical density of $\simeq1\times10^{4}$ cm$^{-3}$,][]{Shirley15} traces the denser parts of the filament including the embedded clumps.

\begin{figure*}
\centering
\includegraphics[width=8.4cm]{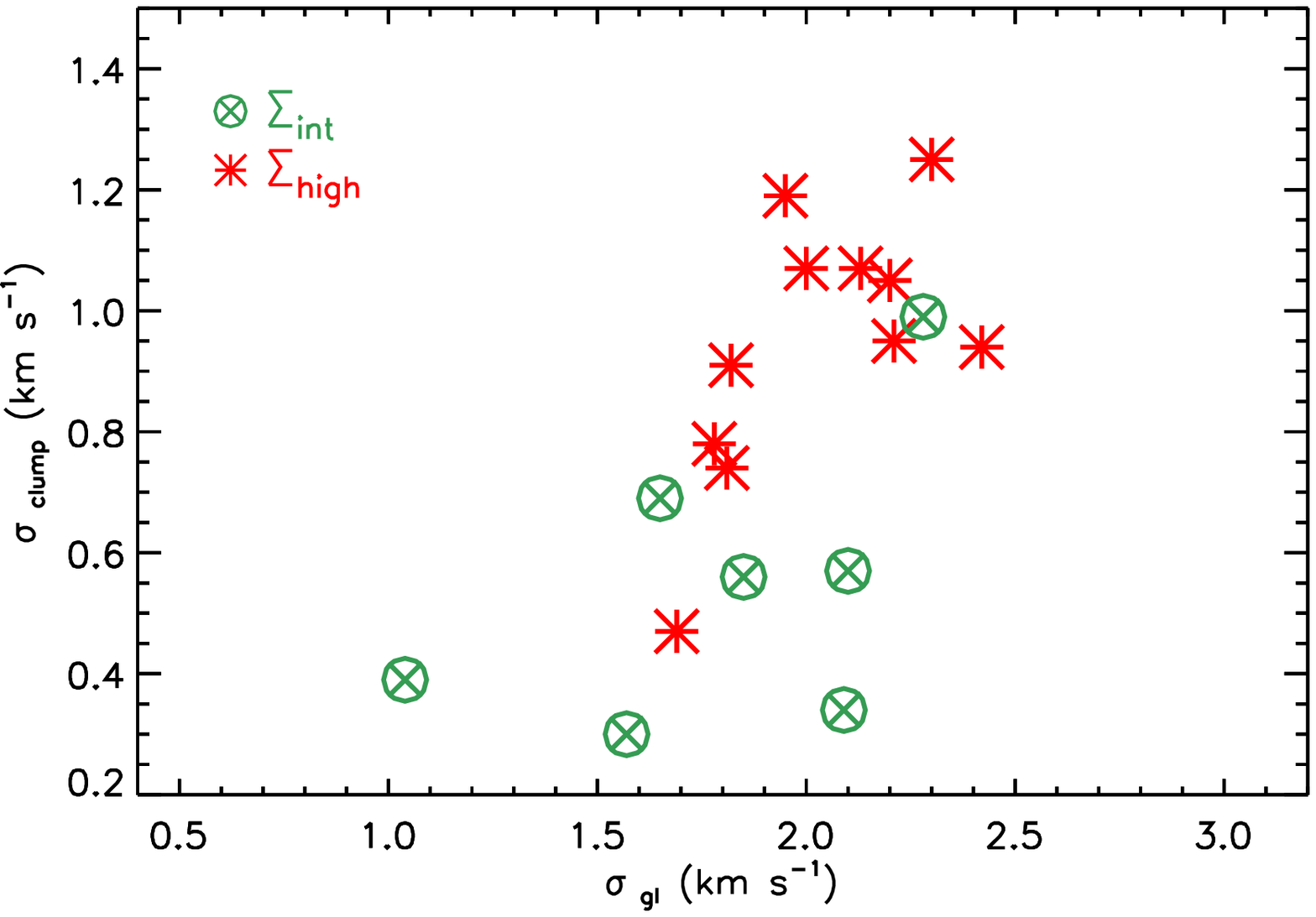} \qquad
\includegraphics[width=8.4cm]{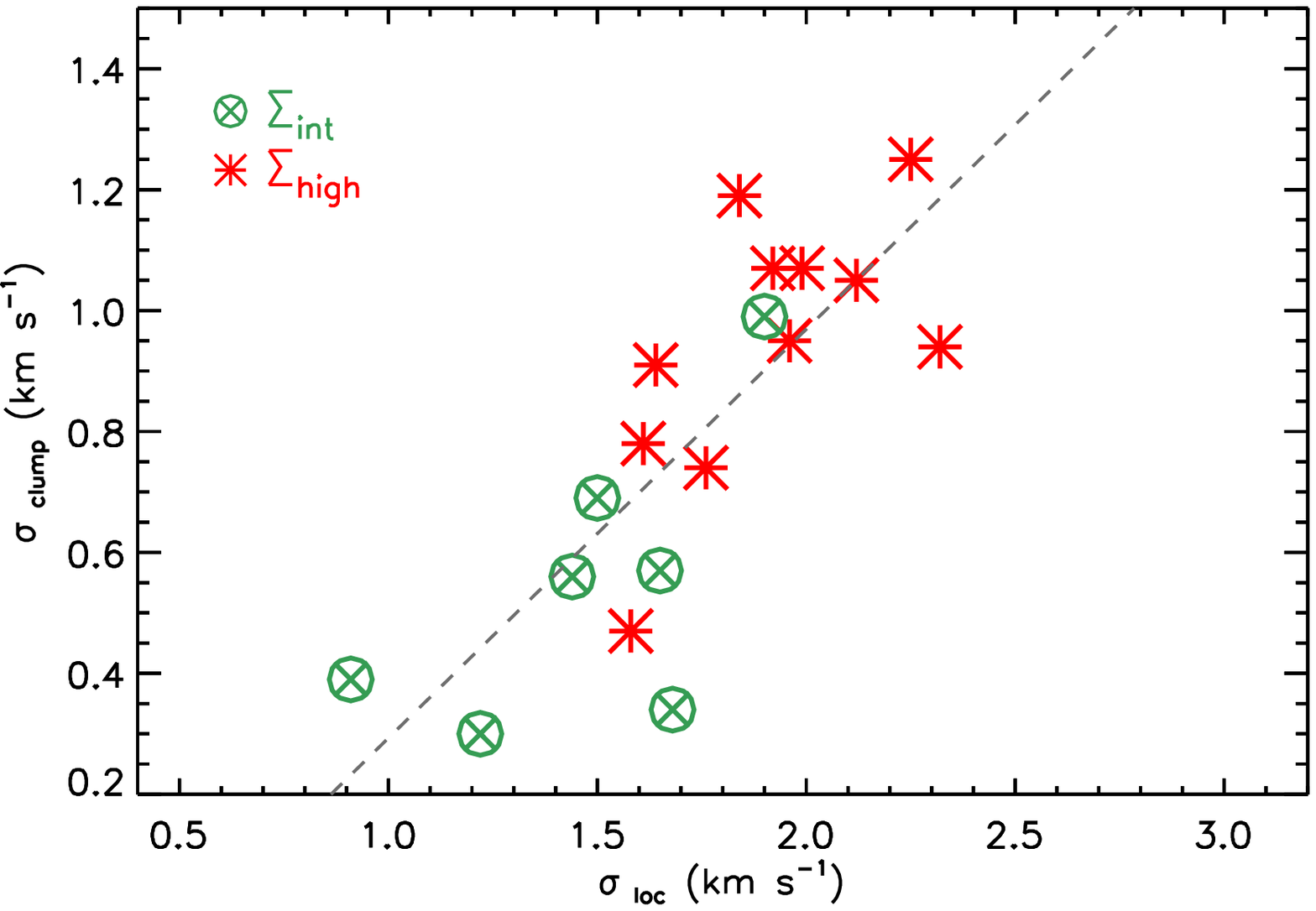}
\includegraphics[width=8.4cm]{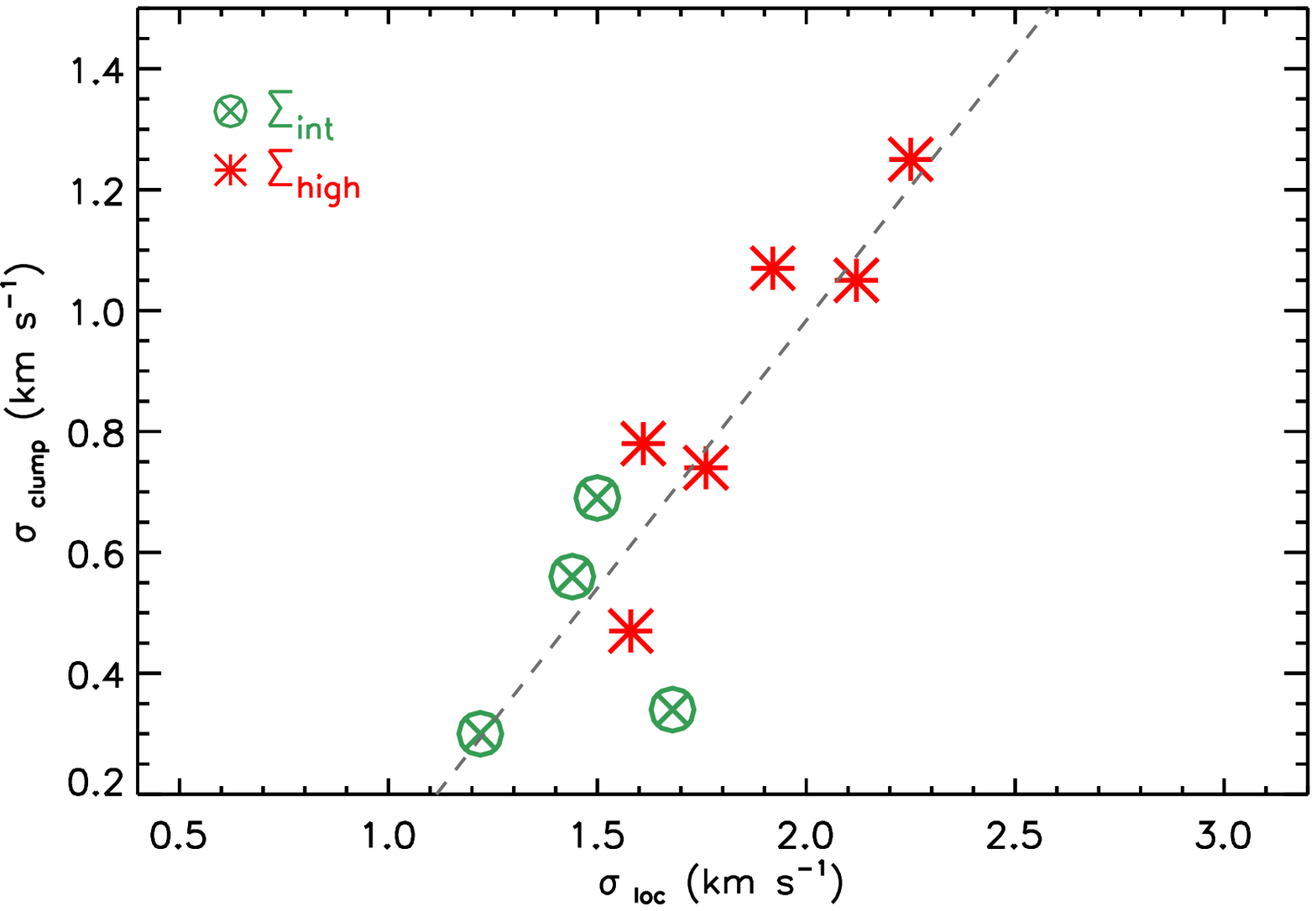} \qquad
\includegraphics[width=8.4cm]{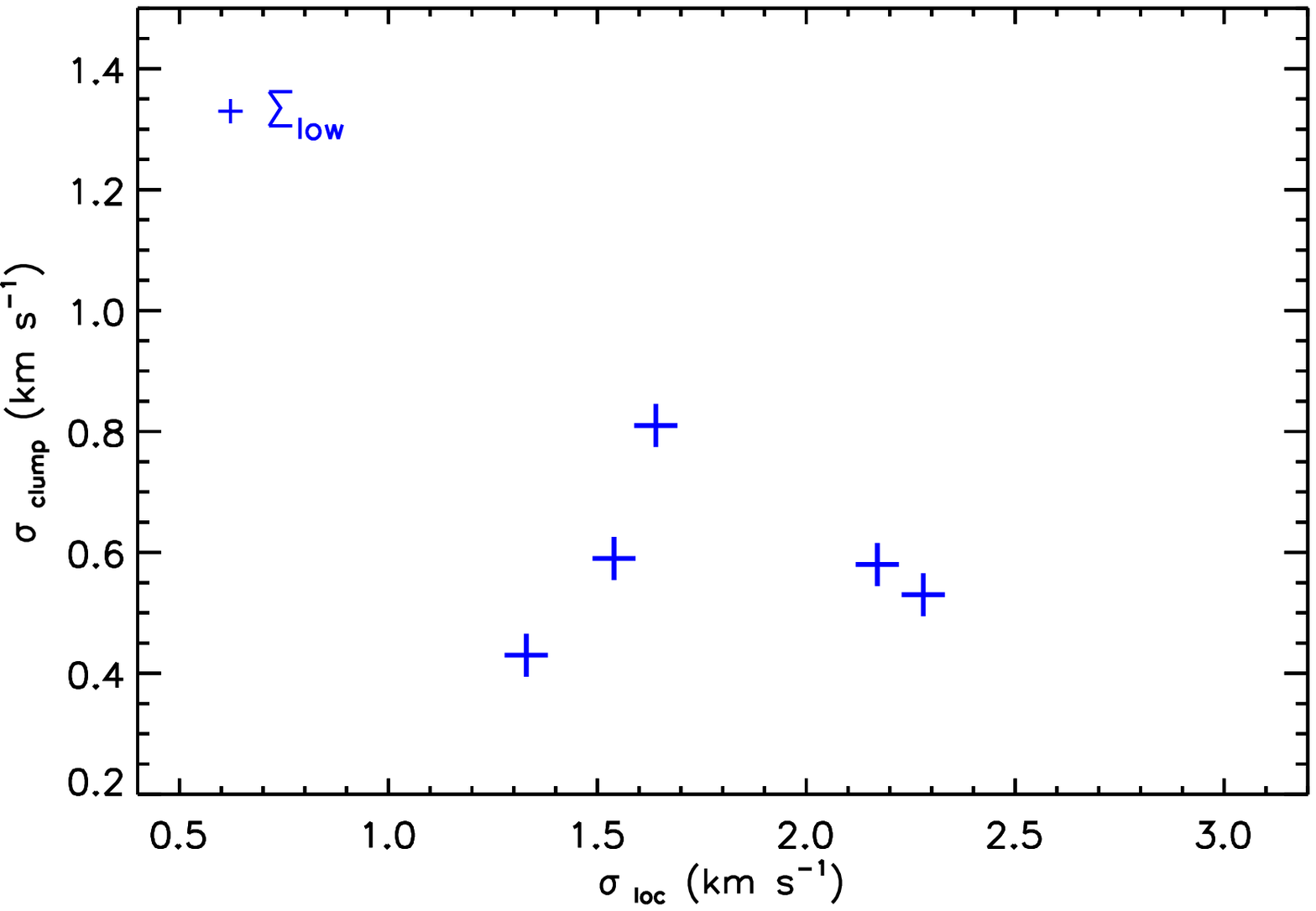}
\caption{Velocity dispersion of the clumps as function of the velocity dispersion of the parent filaments. \textit{Upper-left panel}: clumps and filaments belonging to the $\Sigma_{int}$ and $\Sigma_{high}$ groups. The velocity dispersion of the filaments $\sigma_{gl}$ is estimated including the large-scale gradients. \textit{Upper-right panel}: Same as the previous plot, but using the velocity dispersion of the filaments estimated after the subtraction of the large-scales gradients, $\sigma_{loc}$. The grey-dotted line is the fit of all the sources in log-log space. \textit{Lower-left panel}: same $\sigma_{clump}$ versus $\sigma_{loc}$ plot but including only the 10 regions where the observed clumps are the most massive of each filament. \textit{Lower-right panel}: same relation but for sources belonging to the $\Sigma_{low}$ group. The velocity dispersion of the filament is estimated after the removal of the large-scale gradients, $\sigma_{loc}$.}    
\label{fig:velo_fil_velo_cl}
\end{figure*}

As shown in Figure \ref{fig:velo_fil_velo_cl}, top-left panel, the velocity dispersion of the clumps in $\Sigma_{int}$ and  $\Sigma_{high}$ correlates quite well with the velocity dispersion that includes the large-scales gradients, $\sigma_{gl}$, with a Pearson's coefficient of $\rho_{int}=0.52$ and $\rho_{high}=0.65$ respectively. However, the two distributions occupy slightly different regions of the parameter space. While the velocity dispersion in the $\Sigma_{int}$ clumps is lower than in the $\Sigma_{high}$ clumps, the global non-thermal motions of the parent filaments are very similar, in the range $1.0\leq\sigma_{gl}\leq2.4$ km s$^{-1}$.

In Figure \ref{fig:velo_fil_velo_cl}, top-right panel, we show the same plot but we consider the non-thermal motions without including the large-scale gradients in the filaments, $\sigma_{loc}$. The diagram changes in particular for the $\Sigma_{int}$ regions, because 
the contribution of the large scale term with respect to $\sigma_{loc}$ is, on average, four times higher in this group with respect to the $\Sigma_{high}$ clumps (see Table \ref{tab:filament_params}). As a consequence of the large-scale gradients removal, the correlation for the $\Sigma_{int}$ clumps increases. The two samples together have a significant Pearson's coefficient of $\rho_{int+high}=0.79$, and the correlation has a power-law form $\sigma_{clump}\propto\sigma_{loc}^{0.68}$.

This tight correlation between the kinematics of the filaments and the embedded clumps is confirmed when we plot the same relation but limited to the sample of 10 filaments for which the clumps we observed are the most massive ones (Figure \ref{fig:velo_fil_velo_cl}, bottom-left panel). The relation is stronger, with a very high Pearson's coefficient of 0.86. The correlation has a power-law form $\sigma_{clump}\propto\sigma_{loc}^{0.89}$.

On the other hand, although the statistics is rather poor (only 5 points), the non-thermal motions of the filament appear disentangled from the non-thermal motions of the low-density clumps (Figure \ref{fig:velo_fil_velo_cl}, bottom-right panel) either if we consider the global velocity dispersion value for the filament (with a Pearson's coefficient of $\rho_{low}=0.02$) or $\sigma_{loc}$ (Pearson's coefficient of $\rho_{low}=0.05$), but this is possibly due to the same observational bias discussed before.

\begin{figure}
\centering
\includegraphics[width=8.6cm]{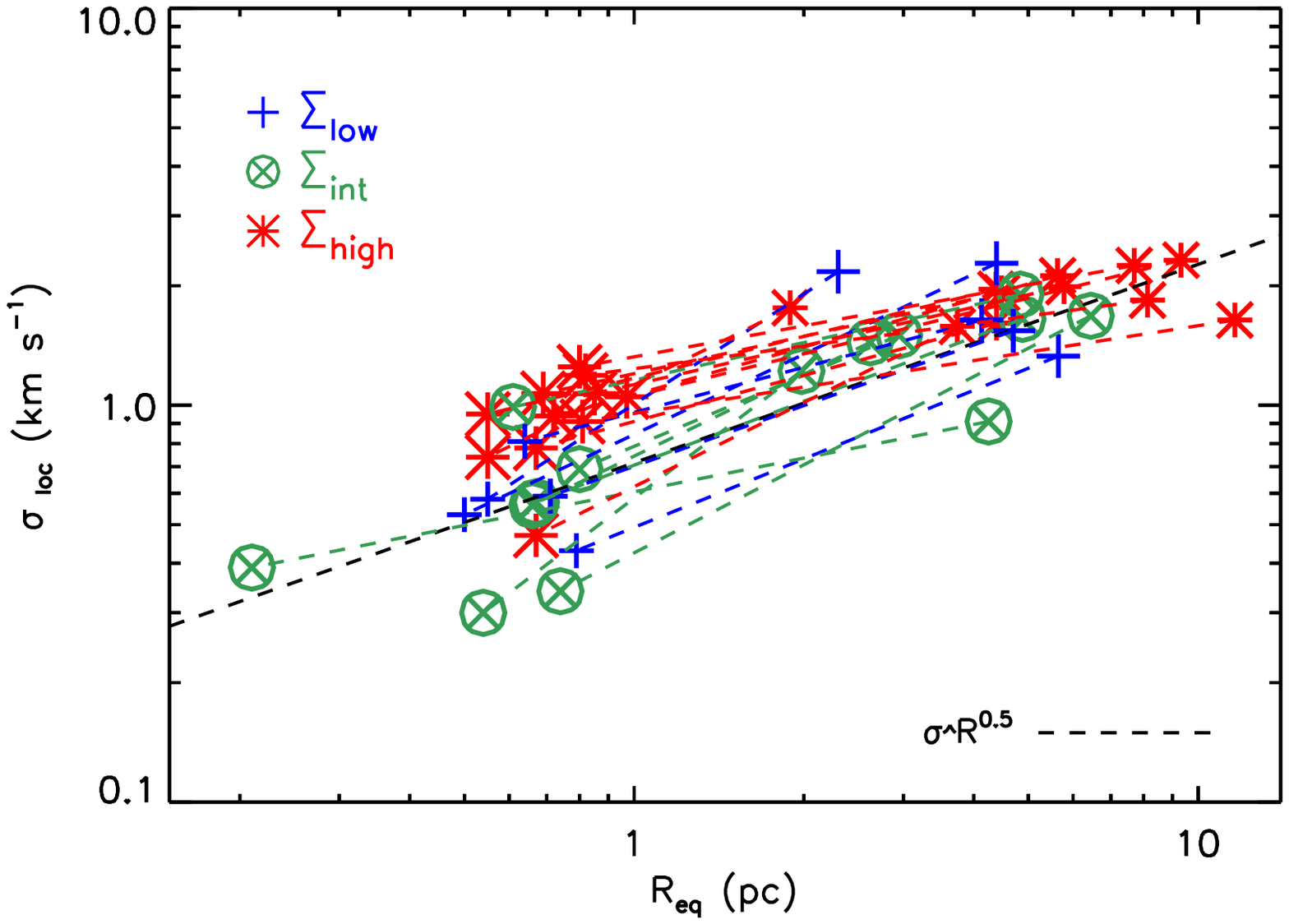}
\includegraphics[width=8.6cm]{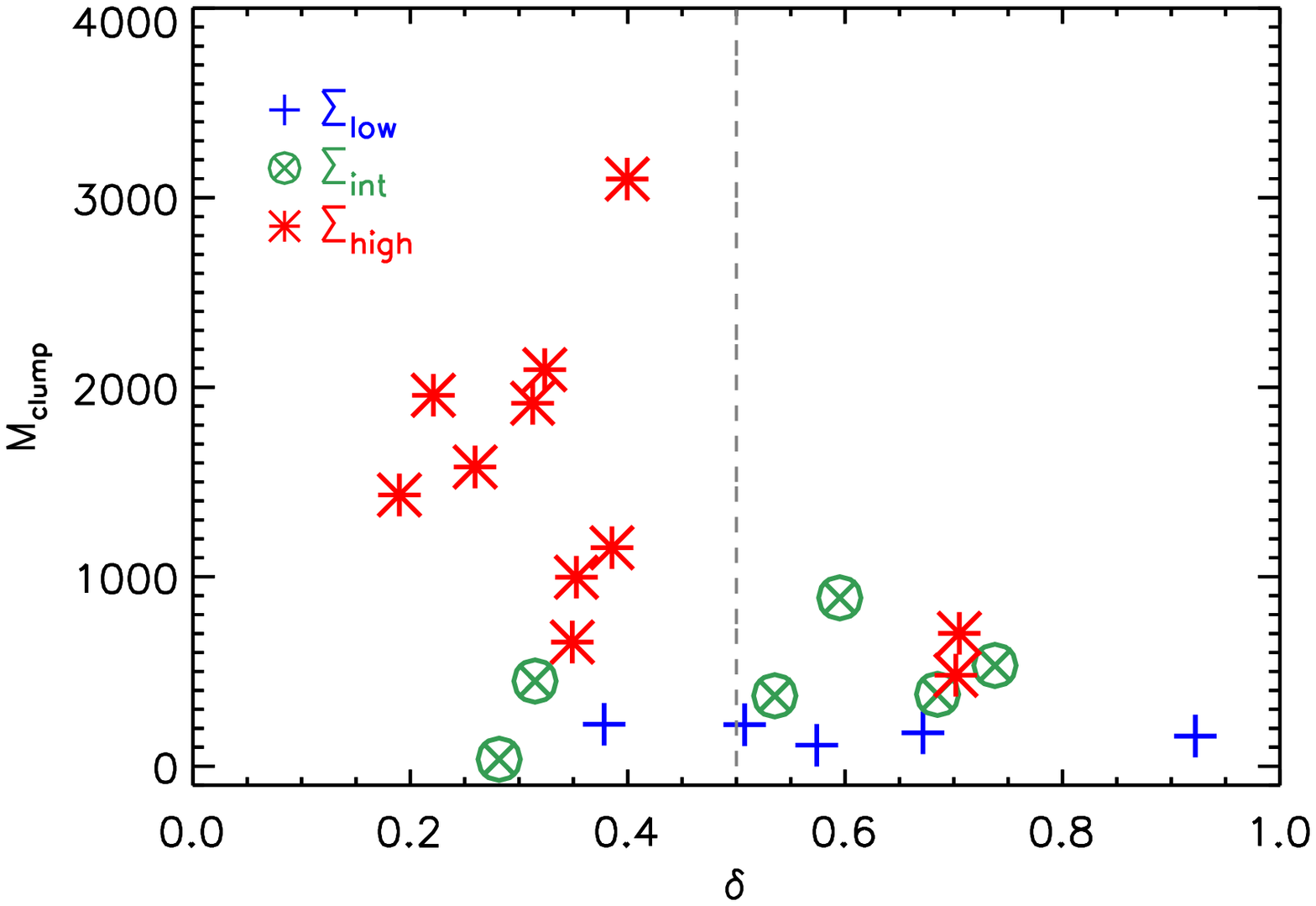}
\caption{\textit{Upper panel}: Larson relation for the clumps and the parent filaments belonging to the three groups. The dotted lines connect each clump (\textit{left points}) with the parent filament (\textit{right points}). The black-dotted line is the $\sigma\propto R^{0.5}$ relation in case of a supersonic turbulent cascade of energy from large down to small scales. \textit{Lower panel}: $M_{clump}$ - $\delta$ relation for our sample. The dashed vertical line is in correspondence of $\delta$=0.5, the fiducial value in case of supersonic turbulent cascade.}

\label{fig:Larson_first_clumps_filaments}
\end{figure}

To further explore the origin of this correlation, in Figure \ref{fig:Larson_first_clumps_filaments}, upper panel we plot again the velocity dispersion - size relation, but this time with dashed lines that connect the values of each clump with the corresponding values of the parent filament (in these plots we consider the equivalent radius \req\ for both clumps and filaments).

Each region now shows an increase of the velocity dispersion moving from the clump to its corresponding filament. If we average the slope of each clump-filament pair to estimate a $\sigma\propto R^{\delta}$ relation for the three groups, we obtain an average value of $\bar{\delta}=0.50\pm0.23$ (the uncertainty is the standard deviation of the distribution), in agreement with the scaling relation expected from a turbulent cascade \citep{McKee07}. The average value of $\delta$ for the three groups separately is $\bar{\delta}_{low}=0.61\pm0.20$, $\bar{\delta}_{int}=0.60\pm0.27$ and $\bar{\delta}_{high}=0.38\pm0.17$, which indicates that the densest clumps have larger internal motions at similar size-scales than the least dense clumps of our sample. To emphasize this trend, in Figure  we report the mass of clump as function of the exponent $\delta$.


We conclude that the dynamics of the filaments and of the embedded clumps are highly correlated.The cascade of turbulence transfers the kinetic energy from the scales of the filaments down to parsec scales within each region. This cascade in each filament produces clumps of various densities, sizes and velocity dispersions imprinted with the kinematics of the parent filament. However, the densest clumps have largest velocity dispersion, the scaling with size indicating that they have excess motions compared with that expected from supersonic turbulence, where $\sigma\propto R^{0.5}$.

\section{Discussion: the interplay between gravity and turbulence at different spatial scales}\label{sec:non_thermal_motions_filaments}
\begin{figure}
\centering
\includegraphics[width=8cm]{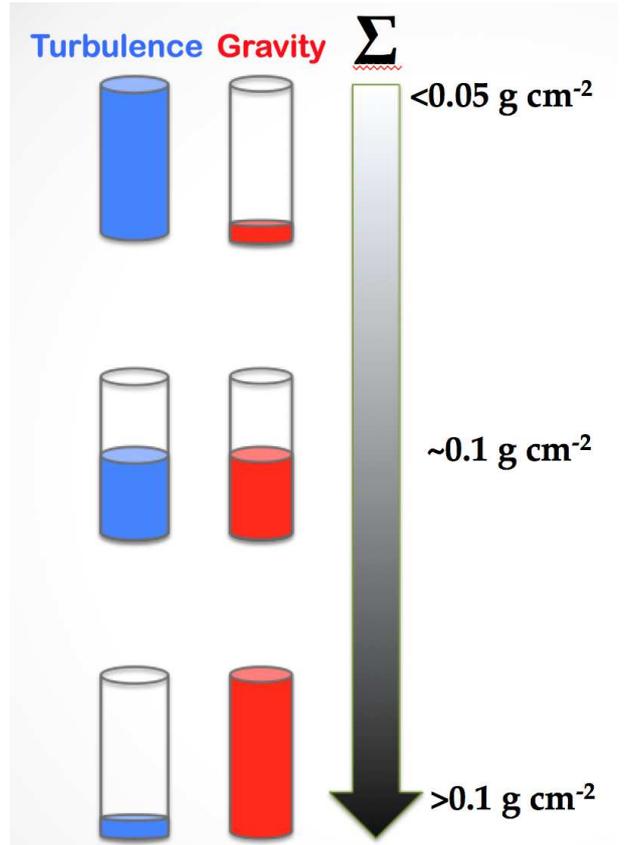}
\caption{A cartoon representation of the interplay between gravity and turbulence in star-forming regions determined from our results. A cloud begins to be compressed due to the effect of the strong ISM turbulence, with minimal contribution from gravity. As the material starts to accumulate in different eddies due to turbulence, these regions increase their surface density and with it the role of gravity in driving the non-thermal motions. As the density reaches the critical value of $\Sigma\simeq0.1$ g cm$^{-2}$, the motions induced by gravity dominate over the turbulence that, from this point, becomes sub-dominant. This critical value of the surface density can be reached at any spatial scales, depending on the initial conditions of the turbulent cloud. Gravity can therefore start to drive the motions at the core, the clump or even the filament scales in star-forming regions embedded in significantly dense clouds.}    
\label{fig:collapse scenario}
\end{figure}

The results discussed in the previous sections suggest that energy is transferred from large down to small scales initially with a continuous cascade of energy, but there is a gradual variation of the dynamics moving from the low-density regime to the high-density one. Figure \ref{fig:collapse scenario} shows a cartoon of this scenario. Turbulence contributes to build-up the material in density perturbations, and once these have reached a critical surface density of $\Sigma\simeq0.1$ g cm$^{-2}$ the effect of gravity becomes apparent in their motions. In the most extreme and dense regions dynamic motions driven by accretion flows are observed at the parsec and larger scales \citep{Peretto13,Traficante17_PI}, suggesting the 0.1 g cm$^{-2}$ surface density is an effective threshold independent of the size scale. If the clouds are embedded in an environment that allows the gas configuration to reach the critical density threshold already at the clump/filament scales, there will be enough material to allow gravity to drive the motions and build-up massive star-forming sites \citep[in agreement with the global collapse model][]{Vazquez-Semadeni19}. In this scenario, the ability to form the most massive stars is already imprinted in the clouds from the very beginning. At the same time, turbulent eddies are likely to be responsible for the formation of the low and intermediate density clumps in all filaments, leaving at the gravity the role to form objects only at the sub-clump size scales. The cascade of turbulence breaks down in any case at the transition regime between supersonic and trans-sonic motions, reached at scales around $\simeq0.1$ pc \citep{Federrath16}.




The lack of relatively isolated high-mass starless cores \citep{Ginsburg12,Motte18} can be a corollary of this scenario: for these objects to be formed, the environment should be very dense. In this case, however, gravity has already overtaken turbulence at parsec-scales, and the whole clump/filament is already in a state of global collapse which triggers the hierarchical fragmentation, as described in the \citet{Vazquez-Semadeni19} model.

\section{Summary}\label{sec:discussion_conclusions}
In this work we have analyzed the dynamics of a sample of 70\mum\ quiet clumps combined with the properties of their parent filaments. The clump analysis has been done on a new sample of 70 \mum\ quiet clumps selected to be in a low-to-intermediate density regime, with $\Sigma\leq0.05$ g cm$^{-2}$, combined with the sample of massive 70\mum\ quiet clumps (with $\Sigma\geq0.05$ g cm$^{-2}$) presented in \citet{Traficante17_PI}. The parent filaments have been extracted from the newly published catalogue of Hi-GAL filaments \citep{Schisano19} and their dynamics analyzed using ancillary CO data.

By dividing the clumps and the parent filaments into three groups, chosen to represent three different surface density regimes for the clumps, we have obtained the following results:

\begin{itemize}
\item[$\bullet$] The kinematics of intermediate and high surface density clumps ($\Sigma\geq0.05$ g cm$^{-2}$) correlates with their surface density, with an increased level of non-thermal motions for higher $\Sigma$. This correlation is not present in the low surface density clumps ($\Sigma<0.05$ g cm$^{-2}$). Combining the results of the three regimes we identify a critical surface density value, $\Sigma\simeq0.1$ g cm$^{-2}$, above which the velocity dispersion starts to correlate with the surface density. This is a similar threshold above which the clumps are all dynamically active and show evidence of parsec-scales infall motions \citep{Traficante17_PII}.

\item[$\bullet$]  The densest star-forming regions are found predominantly in the densest filaments. Similarly, the clumps with the highest velocity dispersion, $\sigma_{clump}$, are observed in filaments with the highest velocity dispersion, $\sigma_{loc}$ (after correction for the large-scale velocity gradients). These two quantities show a power-law form $\sigma_{clump}\propto\sigma_{loc}^{0.89}$. In general, there is a strong correlation between the kinematics of the clumps and of the parent filaments in regions that embed intermediate and high density objects.

\item[$\bullet$] The velocity dispersion - size scaling relation, when evaluated connecting each filament with its embedded clump, shows an average correlation of the form $\sigma\propto R^{\delta}$ with $\delta=0.50\pm0.23$, in agreement with the prediction for a turbulent cascade. This multi-scale dynamics, with energy transferred from large down to small scales, is observed in all regions. In the densest filaments this cascade may be driven by the accretion itself at all scales.

\item[$\bullet$] On similar size scales, the velocity dispersion is on average higher in the densest clumps. If the velocity dispersion of the low density clumps traces the effect of the pure turbulent cascade from larger scales, then the densest clumps have an additional source of internal motions. This would be consistent with gravity playing an enhanced role in these most extreme regions, and would support the global collapse scenario towards extremely massive clumps like e.g. SDC335 \citep{Peretto13}.

\end{itemize}

In conclusion, we observe a multi-scale dynamics within filaments and clumps which we interpret as a continuous interplay between turbulence and gravity. The former drives the non-thermal motions from filaments down to small scales, where gravity begins to dominate the dynamics once the regions reach a surface density above a critical value of $\Sigma\simeq0.1$ g cm$^{-2}$. This value is reached at the clump (or larger) scales within the densest filaments observed in our sample.

Finally, it should be noted that this analysis does not consider the role of the magnetic fields, which could be significant at any spatial scales which require further investigation to fully understand the nature of the observed non-thermal motions as a function of size and density.

\section*{acknowledgements}
This work has been funded by VIALACTEA, a Collaborative Project under FrameworkProgramme 7 of the European Union (Contract n. 607380),that is hereby acknowledged. ADC acknowledge support from the Science and Technology Facilities Council (under grant ST/N00706/1). This work is based on observations carried out under project number 133-15 with the IRAM 30m telescope. IRAM is supported by INSU/CNRS (France), MPG (Germany) and IGN (Spain).

\bibliographystyle{mnras}
\bibliography{Draft_final_arxiv.bbl}

\end{document}